\documentclass[11pt,oneside]{article}

\usepackage[numbers,sort&compress]{natbib}
\setcitestyle{numbers}

\usepackage{rotating}

\usepackage{ragged2e}
\usepackage{placeins}
\usepackage{pbox}

\usepackage{bm}
\usepackage{makecell}

\usepackage[table]{xcolor}

\newcommand{\matrva}[1]{\bm{#1}} 

\newcommand{\E}{\mathbb{E}}
\newcommand{\R}{\mathbb{R}}

\newcommand{\bmu}{\matrva{\mu}}
\newcommand{\bSigma}{\matrva{\Sigma}}
\newcommand{\bw}{\matrva{w}}
\newcommand{\bW}{\matrva{W}}
\newcommand{\br}{\matrva{r}}
\newcommand{\bR}{\matrva{R}}
\newcommand{\bp}{\matrva{p}}
\newcommand{\bP}{\matrva{P}}
\newcommand{\bq}{\matrva{q}}
\newcommand{\bz}{\matrva{0}}
\newcommand{\bOmega}{\matrva{\Omega}}
\newcommand{\btheta}{\matrva{\theta}}

\newcommand{\bX}{\matrva{X}}
\newcommand{\bef}{\matrva{f}}
\newcommand{\beps}{\matrva{\epsilon}}
\newcommand{\bD}{\matrva{D}}
\newcommand{\bmm}{\matrva{m}}
\newcommand{\bmuf}{\matrva{\mu_f}}
\newcommand{\bOne}{\matrva{1}}

\newcommand{\bL}{\matrva{L}}
\newcommand{\bF}{\matrva{F}}
\newcommand{\bV}{\matrva{V}}
\newcommand{\bPsi}{\matrva{\Psi}}
\newcommand{\bI}{\matrva{I}}
\newcommand{\bgamma}{\matrva{\gamma}}
\newcommand{\bGamma}{\matrva{\Gamma}}
\newcommand{\bzee}{\matrva{z}}

\newcommand{\bxi}{\matrva{\xi}}
\newcommand{\bLambda}{\matrva{\Lambda}}
\newcommand{\bDelta}{\matrva{\Delta}}

\newcommand{\bh}{\matrva{h}}
\newcommand{\bx}{\matrva{x}}
\newcommand{\bbeta}{\matrva{\eta}}
\newcommand{\bH}{\matrva{H}}
\newcommand{\bnu}{\matrva{\nu}}

\usepackage{amsthm}
\newtheorem{lemma}{Lemma}

\usepackage{courier}

\usepackage{pdflscape}

\usepackage[hang,flushmargin]{footmisc}

\usepackage{amsmath, amssymb, amsthm, amsfonts, graphicx, mathrsfs, array, stackrel, enumerate, enumitem}
\usepackage{url, overpic, times, eurosym}
\usepackage{booktabs, multirow}
\usepackage{caption,newfloat}
\DeclareCaptionType{InfoBox}

\usepackage{todonotes}

\usepackage{tabularx}
\newcolumntype{Y}{>{\centering\arraybackslash}X}

\usepackage{float}
\restylefloat{table}

\usepackage{textcomp}
\usepackage{amsmath}
\usepackage{array, multirow, bigdelim, makecell, booktabs}



\allowdisplaybreaks

\title{View fusion vis-à-vis a Bayesian interpretation of Black-Litterman for portfolio allocation.} 

\author{Trent Spears\thanks{University of Oxford, Oxford-Man Institute of Quantitative
    Finance.} \thanks{
    Corresponding author.  E-mail: trent@robots.ox.ac.uk}, Stefan Zohren$^*$, Stephen Roberts$^*$
}

\date{\today}

\begin{document}
\thispagestyle{plain} \maketitle

\renewcommand{\abstractname}{Summary}
\begin{abstract}
\noindent The Black-Litterman model extends the framework of the Markowitz Modern Portfolio Theory to incorporate investor views.  We consider a case where multiple view estimates, including uncertainties, are given for the same underlying subset of assets at a point in time.  This motivates our consideration of data fusion techniques for combining information from multiple sources.  In particular, we consider consistency-based methods that yield fused view and uncertainty pairs; such methods are not common to the quantitative finance literature.   We show a relevant, modern case of incorporating machine learning model-derived view and uncertainty estimates, and the impact on portfolio allocation, with an example subsuming Arbitrage Pricing Theory.  Hence we show the value of the Black-Litterman model in combination with information fusion and artificial intelligence-grounded prediction methods. \\

\noindent {\bf Keywords:}  Black-Litterman portfolio allocation, information fusion, machine learning, financial time-series analysis.

\end{abstract}

\section{Introduction}

The Black-Litterman model extends the Markowitz portfolio optimisation framework to incorporate investor beliefs about asset returns \cite{Black91}.  Such beliefs are termed `views'.  In its original formulation, the Black-Litterman framework is such that a given view induces an update of a prior assumption about the mean parameter of a Gaussian return distribution.  Making use of views in this way can result in non-trivial differences to the estimated weights of a portfolio allocation strategy, and hence the relative investment outcome.

\vspace{0.25 \baselineskip}

\noindent Views can be obtained from diverse sources, including market pundit opinions, Reserve Bank statements, and as output of statistical models.  Consequently, in practice, a portfolio manager or trader may possess multiple views about the same underlying asset, or subset of assets, over the next investment period.  A contribution of this work is to both consider and extend the Black-Litterman framework to incorporate multiple such views.

\vspace{0.25 \baselineskip}

\noindent It is natural to formalise elements of the Black-Litterman model with respect to ideas from Bayesian statistics.  See \cite{Jay14} for a review summary, with \cite{Sat00, Kolm17}  intriguing supplements.  We note that the modelling of this paper presupposes the canonical Bayes-based Black-Litterman formalism.  This approach is useful for practitioners in allowing for a strong theoretical grounding of the model prior in the Capital Asset Pricing Model (CAPM) \cite{Sharpe64}, and for the relative ease with which one can incorporate market views.  Specifically, views are given as a function of asset mean return estimates, and accompanied by an uncertainty estimate that encapsulates a notion of confidence about that mean.  The specification of the view uncertainty is a domain for debate in the literature.  In contrast, the uncertainty specification is of critical importance to the mechanics of the Bayesian estimation framework.  The literature often tacitly assumes that uncertainty is to be specified in excess of the given mean estimate.  A prevailing method is to set the uncertainty term in proportion to a quantity such as an estimate of the underlying return covariance.  The proportionality constant can then be tuned akin to a model hyperparameter.  Such ad hoc specification can be unsatisfying; fortunately, scientific modelling offers an alternative.

\vspace{0.25 \baselineskip}

\noindent Machine learning methods are enjoying a resurgence as oracles for financial markets, particularly given the advances underlying modern deep learning.  Indeed, such methods have utility for forecasting time-series, and predictions are often naturally partnered with uncertainty estimates.  Such uncertainties are typically categorised in the literature as \emph{aleatoric} or \emph{epistemic}.  Aleatoric uncertainty relates to the modelled random error of the underlying data generating process, while epistemic uncertainty encompasses much more, including error relating to the model specification, which also subsumes uncertainties about parameter estimates.  These two categories of uncertainty are recognised across multiple research domains, and the models contained therein.   In any case, the estimation of views by statistical forecasting models, coupled with the uncertainty estimate, allows for a more principled approach to setting view parameters when modelling with Black-Litterman.  We describe this further in Section 3 below.  This interpretation stands in stark contrast to existing methods -- with one popular technique described above -- that are more ad hoc.  Our approach has the further benefits of preserving the intent of the canonical model, and has utility for automating an aspect of the investment decision-making process.

\vspace{0.25 \baselineskip}

\noindent  Further our extended modelling framework, we consider combining multiple views for expected returns over the next time increment with respect to view correlation that may be non-zero.  Although we consider the single-step case, we are inspired by the relationship between Black-Litterman and a state-space modelling framework as recently described in \cite{Sch17}.  Indeed, at the core of our approach are techniques from the information fusion literature, that allow us to combine views in a principled way.  In Section 3, we present methods for fusing views assuming both zero and non-zero correlation between them, and consider a conservative fusion method for sets of views that may be inconsistent with one other (according to some distance metric between distributions).  In their relative simplicity, the principles of these fusion methods are understandable, and the resulting fused estimate are explainable.  Some practitioners may find this advantageous relative to alternative black box methods.  In our empirical work, we find fusion-based methods that improve on of the mean and median of our 3 view-generating models in a statistically meaningful way.  This is interesting to the investment practitioner that cannot know ex-ante which of a set of views might outperform; a fusion-based method may be selected instead.

\vspace{0.25 \baselineskip}

\noindent  To show the utility of data fusion, and in connection with recent work by \cite{Kolm17}, we offer an application of the Black-Litterman model assuming a multi-factor model for asset returns.  Factor models present in finance with various specifications; see \cite{Gig22} for a recent review.  We consider a conditional factor model, as inspired by the Arbitrage Pricing Theory (APT) of Ross \cite{Ross76}.  In Section 2, before we further develop view fusion in Section 3, we expound on the Black-Litterman-APT as a Bayesian hierarchical model; a schematic for the model is given in Figure \ref{fg:C} below.  In Section 2.3 we carefully derive the solution for the optimal portfolio weights under an unconstrained Markowitz portfolio allocation as a further contribution to the literature.  With a desire to extend our empirical work, we offer a preliminary model accounting for transaction costs, with details specified in Section 2.4   This model permits a reasonable proof-of-concept, that could be built upon for industrial application or as future academic work.

\vspace{0.25 \baselineskip}

\noindent Notes for the data set construction are offered in Section 4.  To the extent the raw data is available in the

\begin{center}
\makebox[\textwidth]{
  \includegraphics[height=3.7in, width=6.2in]{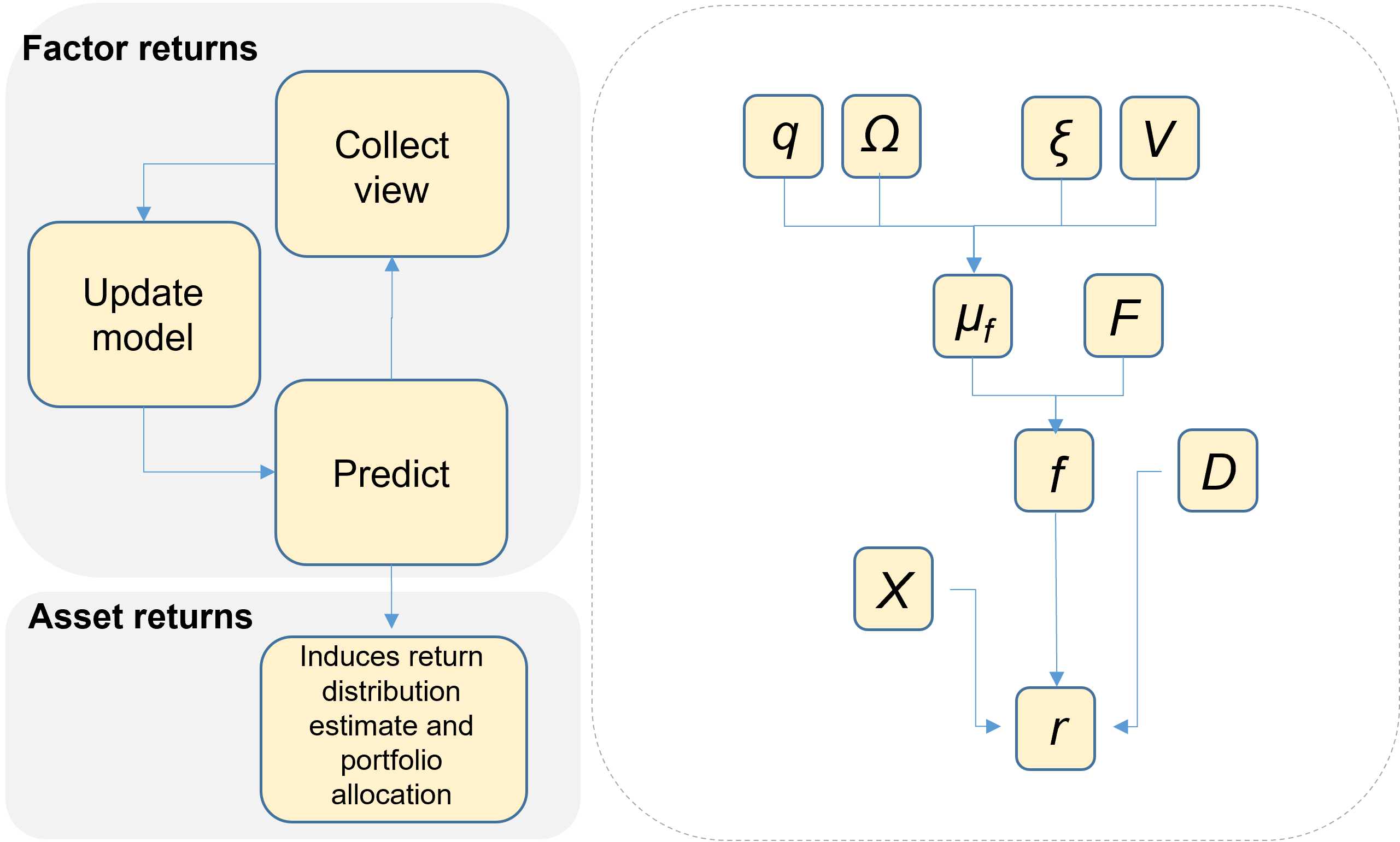}}\par
  \captionof{figure}{LHS: Schematic of the Black-Litterman methodology as it applies to the APT model.  Asset returns are linearly projected onto factor returns, and the algorithm at the level of the factor model is such that at each time-step views are collected, before the factor returns model is updated and predictions gleaned.  RHS: Hierarchical schematic suggestive of the relationship between terms in the BL-APT model, as described in Section 2.
 \label{fg:C}}
\end{center}

\noindent public domain (via the Wharton Research Data Service), these notes could lead to a straightforward replication.  In Section 5, we use the ideas and results of the preceding sections to give empirical results that showcase view fusion whereby estimates -- including measures of uncertainty -- are sourced from a collection of machine learning models.  We discuss the ways in which fusion-based methods lead to investment outperformance relative to methods based on a singular view.  We also constrast our methods with the difficult benchmark of a buy-and-hold investment in the S\&P 500.  We conclude and summarise avenues for future work in Section 6.

\section{Black-Litterman}

\subsection{A recap}

Assume there exist $n$ investable assets within a market with \emph{excess} return vector $\br$ whose distribution is modelled as multivariate normal over discrete time increments, denoted $\br \thicksim \text{MVN}(\bmu, \bSigma)$.  Here $\bmu$ denotes the mean vector of returns, and $\bSigma$ denotes the return covariance.  Then in a Markowitz Modern Portfolio Theory framework \cite{Ma52}, assuming a single-period investment setting, the investor portfolio weights $\bw = (w_1, ..., w_n)$ are a solution to the mean-variance optimisation problem:
\begin{align}
\max_{\bw \in \mathcal{C}} \qquad \E( \bw'\br )  - \frac{\gamma}{2} \text{var}(\bw'\br).  \label{eqMV}
\end{align}
Here $\mathcal{C}$ is a set of weight constraints, if any, and $\gamma>0$ is the investor risk aversion parameter, denoting preferences balancing return and risk.  In the unconstrained case, differentiating eq.\ (\ref{eqMV}), setting it to 0 and solving for the optimal portfolio weights $\bw^*$ in terms of $(\bmu, \bSigma)$ yields 
\begin{align}
\bw^* = (\gamma \bSigma)^{-1} \bmu. \label{eqMVw}
\end{align}
On analysing the second derivative one finds that this solution is indeed a maximum.  In practice, these weights could be solved for -- and a portfolio subsequently implemented -- upon estimation of the return distribution parameters, for example, upon substitution of a computed sample mean vector and sample covariance matrix, given historical data.

\vspace{0.25 \baselineskip}

\noindent  The Black-Litterman model \cite{Black91} extends this framework to incorporate (subjective) investor beliefs about asset returns, that are called `views'.  A single view is assumed to be a function of real-valued vectors of length $n$ denoting portfolio weights $\bp$ and expected asset returns $\bmu$, a real-valued scalar denoting the portfolio value $q$, and a strictly positive real-valued scalar $\epsilon$ denoting uncertainty in the portfolio value, that can be written
\begin{align}
\bp' \bmu = q + \epsilon.  \label{eqViewVec}
\end{align}
To be explicit, let us call a collection of $k$ views a `range', for natural numbers $k$.  Then similar to the original presentation of Black and Litterman, a range of views can be arranged thus:
\begin{align}
\bP \bmu = \bq + \beps, \qquad \beps \thicksim N(\bz, \bOmega). \label{eqViews}
\end{align}
Here $\bP \in \R^{k \times n}$ is a real-valued matrix of portfolio weights, where the $k^{th}$ row is given by $\bp'_k$ as defined above; $\bq \in \R^k$ is a real-valued vector collecting $k$ portfolio values; and $\bOmega \in \R^{k \times k}$ is assumed to be a diagonal matrix of portfolio value estimate uncertainties, where we have further assumed independence between views.

\vspace{0.25 \baselineskip}

\noindent  Given the above formulation, a view can express both  `absolute' and `relative' relationships between mean asset returns.  The case of an absolute (normalised) view corresponds to $\bp$ taking a unit value at the $i^{th}$ entry, and zero otherwise, so that the left-hand side of eq.\ (\ref{eqViewVec}) equals $\mu_i$.  Any other (non-trival) view we call relative; a common relative view, for example, is a belief on the return spread $\mu_i - \mu_j = q$.  This could be expressed by setting alternate positive and negative unit values at the $i$ and $j$ index of $\bp$, and zero elsewhere.  In the work that follows, we assume absolute views.  We further assume that $\bP$ is of full rank.

\vspace{0.25 \baselineskip}

\noindent  Given these preliminaries, the Black-Litterman model is such that the existing return distribution parameter assumption is updated given views.  We understand such updating in the context of Bayesian statistics, as prior authors have noted \cite{Jay14, Sat00, Kolm17, Sch17}.

\vspace{0.25 \baselineskip}

\noindent  The Bayesian specification assumes a prior distribution $p(\btheta)$, where $\btheta$ denotes the target parameter of the return distribution that we wish to express via a parametric probabilistic model.  Most commonly, $\btheta = \bmu$.  The original authors -- and many since -- set the prior given an estimate of the CAPM model \cite{Sharpe64}, and hence underwriting the prior with a foundational theoretical result.

\vspace{0.25 \baselineskip}

\noindent  Next, the view is related to the target parameter via a likelihood function $p(\bq | \btheta )$, such that, by the laws of conditional probability and Bayes' Theorem, an updated posterior distribution for the target parameter is given by
\begin{align}
p(\btheta | \bq ) \propto p(\bq | \btheta) p(\btheta). \label{eqUpdate}
\end{align}
Finally, we can write a posterior predictive distribution for the asset returns given the views as
\begin{align}
p(\br | \bq) = \int p(\br | \btheta) p( \btheta | \bq) d\btheta. \label{eqPredict}
\end{align}
Hence, the expectation and covariance of the random quantity $\br | \bq$ can be computed analytically -- or otherwise estimated if the integral is intractable -- before substitution into, say, eq.\ (\ref{eqMVw}) to yield a solution to the portfolio allocation problem.

\subsection{Black-Litterman in the context of Factor models}

In line with recent work by \cite{Kolm17}, we offer an application of the Black-Litterman model assuming a factor model for asset returns, and we conveniently conform with some of their notation.

\vspace{0.25 \baselineskip}

\noindent For context, we note that in many investment studies of equity markets the empirical estimation of the return covariance is challenging, especially when the universe of equities under consideration is large.  The linear factor model that we outline below, whereby $k \ll n$, is relatively practical and efficient.  This has contributed to its popularity within the literature, and bolsters the case for industrial use \cite{Fab06}.  We consider a conditional factor model, as inspired by the Arbitrage Pricing Theory (APT) of Ross \cite{Ross76}.  In the single-period investment setting, we model
\begin{align}
\br = \bX \bef + \beps_r, \qquad \beps_r \thicksim N(\bz,\bD). \label{eqR}
\end{align}
Here $\br$ is the $n$-dimensional random vector containing the cross-section of excess returns for $n$ equities over the next time increment $[t,t+1]$; $\bX$ is a non-random $n \times k$ matrix of observable factor exposures estimated at $t$, for $k$ the number of explanatory asset factors; $\bD$ is a covariance matrix for the returns, also known at $t$ and assumed to be diagonal; and $\bef$ is a $k$-dimensional latent factor vector that we estimate at each time step by cross-sectional regression.  Within a Black-Litterman framework, consider that our parameter of interest is now $\btheta = \bmuf$, such that 
\begin{align}
\bef = \bmuf+ \beps_f, \qquad \beps_f \thicksim N(\bz,\bF). \label{eqF}  
\end{align}  
We refer to the elements of $\bmuf$ as the factor risk premia.  

\vspace{0.25 \baselineskip}

\noindent Black-Litterman in the context of APT, henceforth labelled `BL-APT', amounts to a Bayesian hierarchical model.  A schematic for the model is given in the right-hand panel of Figure \ref{fg:C}.  We note that in comparison to the Black-Litterman model applied to the return mean, an analogue to the CAPM prior does not exist.  In the work that follows, we set the prior parameters as a simple function of recent data, per the discussion in Section 5.  On the other hand, view estimates for the factor risk premia at time $t$ induce return forecasts (with uncertainties) over the next time period.  These parameter estimates can then be utilised to estimate portfolio weights.

\vspace{0.25 \baselineskip}

\noindent In the next section, we carefully show the derivation for the optimal portfolio weights of an unconstrained `BL-APT' as a reference contribution to the literature.

\subsection{Solving for optimal weights -- Hierarchical Bayesian Black-Litterman for APT}

We can express our model for asset returns, factor returns and views carefully as a hierarchical Bayesian model, with a suggestive schematic shown in Figure \ref{fg:C} above.  We derive the optimal portfolio allocation after finding $p( \br | \bq)$, as defined below.

\vspace{0.25 \baselineskip}

\noindent We begin by writing the joint distribution of random variables of interest as a conditional probability:
\begin{align*}
p( \br, \bef, \btheta, \bq) = p( \br, \bef | \btheta, \bq) p( \btheta, \bq).
\end{align*}
Substituting $p(\btheta, \bq) = p(\btheta | \bq) p(\bq)$ and dividing both sides by $p(\bq)$ yields
\begin{align*}
p( \br, \bef, \btheta | \bq) = p( \br, \bef | \btheta, \bq) p( \btheta | \bq).
\end{align*}
Integrating both sides over $\btheta$ we find that
\begin{align*}
p( \br, \bef | \bq) = \int \underbrace{p( \br, \bef | \btheta, \bq)}_{= p(\br, \bef | \btheta)} p( \btheta | \bq) d\btheta.
\end{align*}
Again integrating both sides, this time with respect to $\bef$, we have that
\begin{align}
p( \br | \bq) = \iint p( \br, \bef | \btheta)  p( \btheta | \bq) d\btheta d\bef. \label{eqTMP}
\end{align}
With one more simplification, writing $p(\br, \bef | \btheta) = p(\br | \bef, \btheta) p(\bef|\btheta) = p(\br | \bef) p(\bef |\btheta)$, we find
\begin{align}
p( \br | \bq) = \int p(\br | \bef) \underbrace{\int p( \bef | \btheta)  p( \btheta | \bq) d\btheta}_{=p(\bef|\bq)} d\bef. \label{eqTMP2}
\end{align}
Since we assume $\btheta = \bmuf$, there holds the well-known facts that the posterior $p(\btheta | \bq)$ is normal and that the posterior predictive $p(\bef | \bq)$ is normal.  Hence it is clear that $p(\br | \bq)$ is also normal, with its expectation and covariance given in closed-form by the following three-step process.

\vspace{0.25 \baselineskip}

\noindent Firstly, choosing the prior $p(\btheta) \thicksim N( \bxi, \bV)$, by Bayes' rule for linear Gaussian systems, as in \cite{Murphy12}, we have that the posterior $p(\btheta | \bq)$ is multivariate normal with covariance and mean 
\begin{align*}
\text{var}(\btheta | \bq) &= (\bV^{-1} + \bOmega^{-1})^{-1}, \\
\E(\btheta | \bq) &= \text{var}(\btheta | \bq) \cdot (\bV^{-1} \bxi + \bOmega^{-1} \bq).
\end{align*}
Secondly, recall the convolution integral for sums of random variables: if $\bX_i \stackrel{i.i.d.}{\thicksim} N(\bmu_i, \bSigma_i), i=1,2$, then
\begin{align}
f(\bzee) := \int f_{\bX_1}(\bx) f_{\bX_2}(\bzee-\bx) d\bx  \thicksim N(\bmu_1+\bmu_2, \bSigma_1 + \bSigma_2 ). \label{eqConv}
\end{align}
Writing the inner integral of eq.\ (\ref{eqTMP2}) with the change of variable $\bef \mapsto \bef - \btheta$, we find
$$
p(\bef | \bq) = \int \underbrace{p( \bef - \btheta | \btheta)}_{\thicksim N(\bz, \bF)}  p( \btheta | \bq) d\theta
$$
so that by eq.\ (\ref{eqConv}) we have that $p(\bef | \bq)$ is normally distributed with expectation and covariance given by
\begin{align*}
\E(\bef | \bq) &= (\bV^{-1} + \bOmega^{-1})^{-1} \cdot (\bV^{-1} \bxi + \bOmega^{-1} \bq) = \E(\btheta | \bq), \\
\text{var}(\bef | \bq) &= (\bV^{-1} + \bOmega^{-1})^{-1} + \bF = \text{var}(\btheta | \bq)  +\bF.
\end{align*}
Thirdly, we solve the outer integral of eq.\ (\ref{eqTMP2}) (deferring some of the details of the algebraic manipulations to the Appendix, for brevity).  Writing
\begin{align}
p( \br | \bq) = \int p(\br | \bef) p( \bef | \bq) d\bef \label{eqTMP3},
\end{align}
we collect and expand the terms in the exponent up to a factor of $-\frac{1}{2}$, complete the square in $\bef$, and integrate what is recognisable as a Gaussian integral.  By the recursive nature of the integrals, it is clear that we also retrieve a Gaussian for $\br | \bq$.  Hence, as supported by further calculations within the Appendix below, we have that the posterior predictive distribution $p (\br | \bq)$ is multivariate normal with covariance and mean 
\begin{align*}
\text{var}(\br | \bq) &= \big[ \bD^{-1} - \bD^{-1} \bX \big[\bX^T \bD^{-1} \bX + \text{var}(\bef|\bq)^{-1}\big]^{-1} \bX^T \bD^{-1} \big]^{-1}, \\
\E(\br| \bq) &= \text{var}( \br | \bq ) \bD^{-1} \bX \big[\bX^T \bD^{-1} \bX + \text{var}(\bef|\bq)^{-1}\big]^{-1} \text{var}( \bef | \bq)^{-1} \E(\bef | \bq).
\end{align*}
By eq.\ (\ref{eqMVw}), we write the optimal weights for a single-step Black-Litterman portfolio allocation as
\begin{align}
\bh_{blb}^* &=\gamma^{-1} \text{var}(\br | \bq)^{-1} \E(\br| \bq) \nonumber \\
&=\gamma^{-1} \bD^{-1} \bX \big[\bX^T \bD^{-1} \bX + \text{var}(\bef|\bq)^{-1}\big]^{-1} \text{var}( \bef | \bq)^{-1} \E(\bef | \bq). \label{eqBLB}
\end{align}

\vspace{0.25 \baselineskip}

\noindent This expression differs meaningfully with other optimal weights given in the literature.  This is the case in that we have carefully accounted for all of the specified terms within the hierarchy of the returns model.

\subsection{An optimal weight model under transaction costs}

\noindent  In this section we consider a more realistic objective function for solving for portfolio weights, in that it incorporates transaction costs.  We do this by making a minimal adjustment to the Markowitz objective of eq.\ (\ref{eqMV}):
\begin{align}
\max_{\bw \in \mathcal{C}} \qquad \E( \bw'\br ) - \frac{\gamma}{2} \text{var}(\bw'\br) - TC,  \label{eqMV2}
\end{align}
with $TC$ denoting transaction costs measured in percentage of invested wealth.  Of the many ways to account for transaction costs, we broadly follow the single-period model recently presented in \cite{Jen22}.  Despite the differences in our application domains, our analysis and results contrast somewhat consistently, and certainly interestingly, with theirs.  

\vspace{0.25 \baselineskip}

\noindent Firstly, we choose a transaction cost model inspired by \cite{Gar13} so that  transaction costs measured in dollars, $TC^d$, is given by
$$
TC_t^d = \frac{1}{2} \bDelta_{t} ' \bLambda_t \bDelta_{t}.
$$
Here $\bDelta_{t}$ is the portfolio turnover vector for a time increment, defined as
$$
\bDelta_{t} := \Pi_t (\bw_t-\bR_{t-1} \bw_{t-1}),
$$
and the `market impact' vector $\bmm$ is given by
$$
\bmm_t = \frac{1}{2} \bLambda_t \bDelta_{t}.
$$
We use the subscript $t$ to explicitly account for time-dependence.  The real scalar $\Pi_t$ denotes total wealth investable at time $t$. Clearly, given the above specification dollar transaction-costs erode dollar returns quadratically with increasing wealth.  The term $\bR_{t-1}$ denotes a diagonal matrix whose diagonal elements account for the realized returns for the assets held in the portfolio at $t-1$, over the time step $[t-1,t)$, and adjusted for wealth delta.  We assume that the term $\bLambda_t$ is a diagonal matrix whose elements depend on the daily (dollar) volume traded in each underlying asset.  Given the recent estimates of \cite{Fraz18}, we assume a 10 basis point market impact when trading 1\% of the daily dollar volume $\bL_t$ of US equity underlyings whether long \emph{or short}, so
\begin{align}
\bmm_t &= 10 \text{bps} \cdot \bOne  = \frac{1}{2} \bLambda_t  \times 1\%  \bL_t \nonumber \\
&\implies \bLambda_t =  \frac{1}{5} \bL_t^{-1}.  \label{LambdaV}
\end{align}
In the work that follows, we set the values for daily dollar volumes at their 6 month rolling average.  Further, we assume that investor wealth grows proportionally with the market; specifically, we make the somewhat arbitrary assumption that at each time step the total investable capital is one-tenth of the sum of the daily dollar volumes of the assets in our trade universe, as known at the most recent time step.
  
\vspace{0.25 \baselineskip}

\noindent Hence eq.\ (\ref{eqMV2}) can be written, noting that $TC = TC^d/\Pi$, as
\begin{align}
\max_{\bw_t \in \mathcal{C}} \qquad \E\big( \bw_t'\br_t \big)  - \frac{\gamma}{2} \text{var}\big(\bw_t'\br_t\big) - \frac{\Pi_t}{2}\big(\bw_t-\bR_{t-1} \bw^*_{t-1}\big)'\bLambda_t \big(\bw_t-\bR_{t-1} \bw^*_{t-1}\big)  \label{eqMV3}
\end{align}
and conveniently solved in closed-form for optimal weights:
\begin{align}
\bw_t^* = \big(\gamma \bSigma_t + \Pi_t \bLambda_t\big)^{-1}\big(\bmu_t + \Pi_t \bLambda_t \bR_{t-1} \bw^*_{t-1}\big). \label{eqMVw2}
\end{align}

\noindent Hence the optimal portfolio weights for the BL-APT model can be found by substituting estimates for $\E(\br| \bq)$ and $\text{var}(\br | \bq)$ into eq.\ (\ref{eqMVw2}), as an alternative to eq.\ (\ref{eqBLB}), when taking transaction costs into account.

\section{View fusion}

\subsection{Preliminaries}

\noindent The Black-Litterman model -- sans solving for the portfolio weights -- can be expressed via a state-space representation; this was recently presented in \cite{Sch17} in the context of modelling temporal dependency in asset returns and parameter view estimates\footnote{There exists an extensive literature on state-space modelling, with applications across many domains.  See, for example, \cite{Date10} for an instructive review of filtering methods for mathematical finance.}.   In a simple case, consider ranges of views generated from a source at discrete time increments $k=0,1,...$ .  We can infer the underlying target mean $\bmu_k$ in an online way, via the linear time-varying system of equations:
\begin{align}
\bmu_{k+1} &= \bmu_k + \beps^{\bmu}_k, \label{eqS} \\
\bq_{k} &= \bP_k \bmu_k + \beps^{\bq}_k. \label{eqM}
\end{align}
Here $\beps^{\bmu}_k$ and $\beps^{\bq}_k$ model independent zero-mean white-noise processes with respective covariances $\bPsi_k$ and $\bSigma_k$.  We call equations (\ref{eqS}) and (\ref{eqM}) the `state' and `measurement' equations, respectively.  Solving the state-space representation of Black-Litterman, assuming absolute views, we have the probability distribution for $\bmu_k | \bq_k$, analogous to eq.\ (\ref{eqUpdate}) and solved equivalently to solving for $p(\btheta|\bq)$ per Section 2.1.  The predictive distribution for $\bmu_{k+1} | \bq_k$, analogous to eq.\ (\ref{eqPredict}) can be solved equivalently to solving for $p(\bef|\bq)$, also of Section 2.1.   On the other hand, within the Black-Litterman literature, as in our model from Section 2, the prior is typically re-initialised to $p(\bmu_0)$ at each time-step, rather than dynamically updated given a sequence of view estimates.  In the following work, we follow this approach, and leave consideration of prior updates based on a larger subset of recent data as future work.  In any case, such state-space modelling naturally inspires an extension of the Black-Litterman framework to handle the combination -- or `fusion' -- of multiple views held about a particular portfolio.

\subsection{Information fusion}

To this point we have described the Black-Litterman model assuming a singular source for a range of views.  Next, we consider the realistic -- though in the context of Black-Litterman modelling, unexplored -- case of multiple sources generating views simultaneously for each time increment, and analyse the impact on the predictive distribution.  We extend eq.\ (\ref{eqM}) thus (for brevity, keeping to the case of absolute views):
\begin{align}
\bq_{k,s} &= \bmu_{k,s} + \beps^{(\bq)}_{k,s}, \label{eqMs}
\end{align}
where $\bq_{k,s}$ denotes the view is from source $s$, $ 1 \le s \le S$, and each $\beps^{(\bq)}_{k,s}$ is a zero-mean white-noise process with covariance $\bSigma_{k,s}$.  Multiple views could realistically arise in the context of implementing a systematic trading or asset management strategy, whereby multiple predictions across multiple (potentially, broadly diverse) models need to be fused before capital allocation.  

\vspace{0.25 \baselineskip}

\noindent Information fusion has been common to the state-space literature since the work of \cite{Wil76}.  However, fusion is not limited to the state-space framework, and instead applies more generally to the problem of fusing multiple estimates of an underlying random variable\footnote{Various approaches exist in the data fusion literature for incorporating information from multiple sources; introductory reviews can be found in \cite{Kha11, Cas13}.}.  Correspondingly, data fusion can take place at the level of the state or measurement equations, though we assume the former for our empirical work below.  

\vspace{0.25 \baselineskip}

\noindent Typically, fusion techniques amount to solving for an optimal fused estimate $ (\widehat{\bmu}, \widehat{\bSigma})$ for the true underlying values $ (\bmu, \bSigma)$ given a collection of $S$ underlying estimates $\{ (\widehat{\bmu}_s, \widehat{\bSigma}_s): 1 \le s \le S \}$, with the mean 

\begin{center}
\makebox[\textwidth]{
  \includegraphics[height=3.9in, width=5.6in]{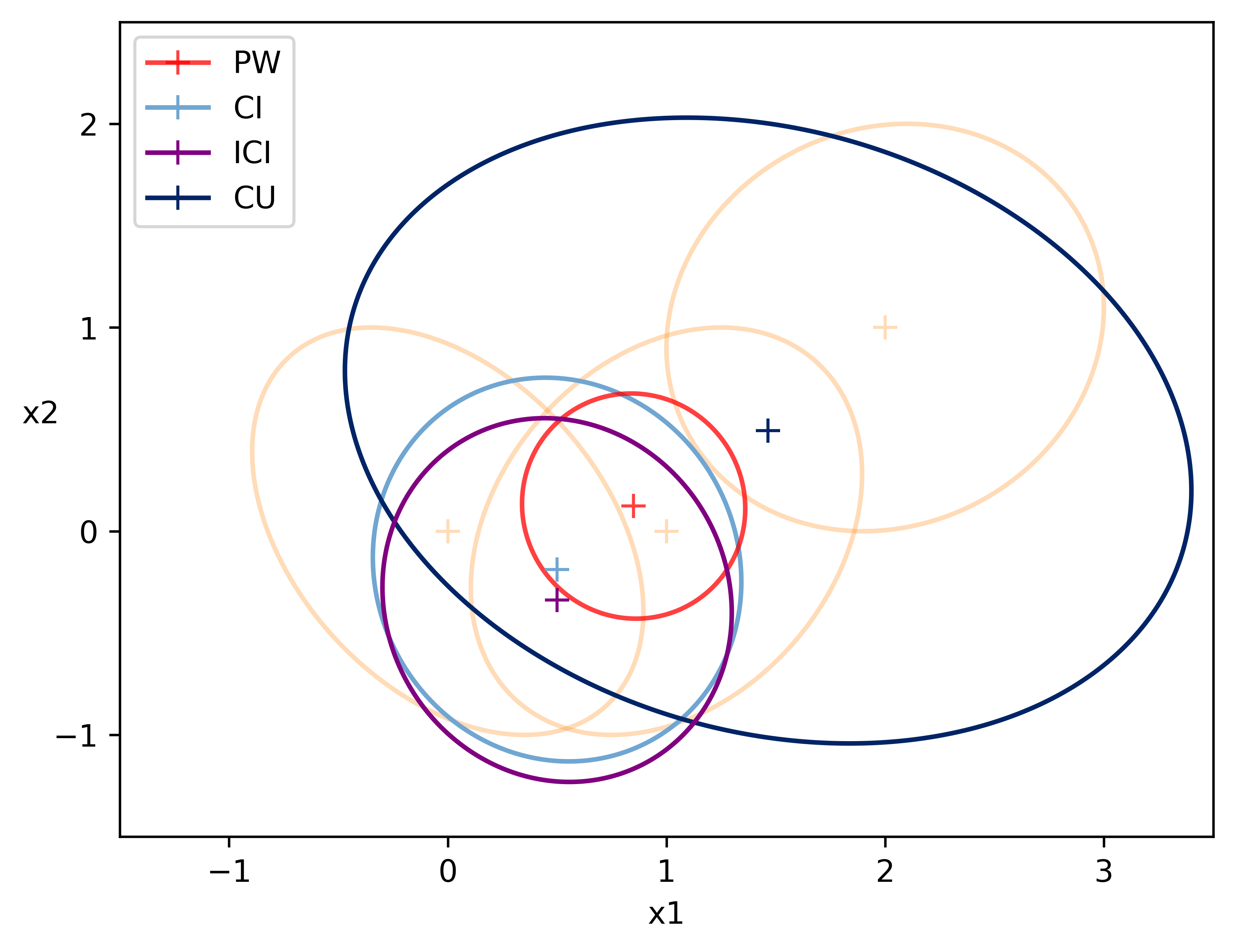}}\par
  \captionof{figure}{Concentration ellipses for three example bivariate Gaussian measurements of a single state $\bx = (x_1, x_2)$ (light orange), and the optimal fused state estimates for the four methods outlined in Section 3.
 \label{fg:CU}}
\end{center}

\noindent parameter a weighted linear combination of the underlying estimates:
\begin{align}
\widehat{\bmu} = \sum_s  \bW_{s} \text{ } \widehat{\bmu}_{s}. \label{eqLinFuse}
\end{align}

\noindent [We have omitted the subscript $k$, for ease of notation.]  `Optimal' fusion is such that (i) $\widehat{\bSigma}$ is minimal in some sense -- typically with respect to matrix trace or determinant; and also that (ii) $\widehat{\bSigma}$ is \emph{consistent}.  Within the fusion literature, a consistent estimate is such that $\widehat{\bSigma} \ge \bSigma$, in the sense that $\widehat{\bSigma} - \bSigma$ is positive semi-definite.  That is, the consistent estimate subsumes a covariance no less than the covariance $\bSigma$ of the distribution of the true underlying process; see, for example, \cite{Uhl95}.  We note that this definition is independent of the well-known definition of a consistent estimator from the statistical literature, whereby a parameter is called consistent if it converges in probability to the true value of the parameter.  The consistency property is crucial for many investors conscious of risk, in particular those that do not want to understate risk\footnote{In the case of (multi-period) Kelly investing, to which the fusion of this paper could directly apply, investment size is inversely proportional to variance.  But the investor faces almost sure ruin if they sufficiently over-invest (which can happen if uncertainty is estimated to be too small) relative to the optimum strategy  \cite{Thorp06}.}.  Another benefit of consistency is that probabilistic bounds for the state of the system can be deduced, as discussed in \cite{Ree10}.

\vspace{0.25 \baselineskip}

\noindent Given the mean and covariance of a bivariate random variable, recall the `concentration ellipse' \cite{Dem69} defined as the locus of points $\{\bx : (\bx-\bmu)^T \bSigma^{-1} (\bx-\bmu) = 1\}$.  For visual intuition, we depict such ellipses, estimated using the fusion methods described in the following subsections, in Figure \ref{fg:CU} above.  These results are shown for the case of three underlying data sources, assumed to each yield a noisy estimate of the same target.

\subsubsection{Precision-weighted fusion}

We first consider precision-weighted (PW) fusion for the case that the cross-covariances between sources is known to be 0.  Suppose that for each $s$, $\widehat{\bmu}_{s}$ is an unbiased estimate, and each $\widehat{\bSigma}_{s}$ is a known error covariance estimate that is consistent.  It follows that the fused estimate given in eq.\ (\ref{eqLinFuse}) is an unbiased estimate if and only if $\sum_s \bW_{s} = \bI$, for weight matrices $\bW_{s}$, and $\bI$ the identity matrix.  To find the weight estimates we solve
\begin{align*}
\min_{\bW} \text{ trace}(\widehat{\bSigma}) \text{ subject to } \sum_s \bW_{s} = \bI.
\end{align*}
Let $\widehat{\bSigma}^{ij}$ denote the cross-covariance estimate between the unique sources $s_i$ and $s_j$, and write
\begin{align}
\widehat{\bSigma}= \sum_s \bW_{s} \widehat{\bSigma}_{s} \bW_{s}^T + \sum_{s_i \ne s_j} \bW_{s_i} \widehat{\bSigma}^{ij} \bW_{s_j}^T. \label{eqXCov}
\end{align}
Assuming the cross-covariances are everywhere 0, solving for the optimal weights one finds that, on weight substitution,
\begin{align}
\widehat{\bSigma} &= ( \widehat{\bSigma}_{1}^{-1} + ... + \widehat{\bSigma}_{S}^{-1})^{-1}, \label{fuseKC1} \\ 
\widehat{\bmu} &= \widehat{\bSigma} ( \widehat{\bSigma}_{1}^{-1} \widehat{\bmu}_{1} + ... + \widehat{\bSigma}_{S}^{-1} \widehat{\bmu}_{S}). \label{fuseKC2}
\end{align}

\noindent Further details of this well-known result can be found in \cite{May79, Wil76}.  The analogous result for fusing two sources that have known, non-zero correlation is given in \cite{YBS81}, though for our financial modelling framework this is less useful since the cross-correlations are not estimated.  On the other hand, the case of unknown, potentially non-zero cross-correlation is the domain of the Covariance Intersection method, as presented in the following subsection.

\subsubsection{Covariance Intersection}

One could reasonably expect that views on the same underlying asset returns, as functions of similar or equal underlying data, exhibit non-zero cross-correlation.    However, the cross-correlation is not necessarily known, or easily estimable.  Such a setting is the domain of application for the Covariance Intersection (CI) fusion method of \cite{Uhl97}.  The insight of this approach is that, given eq.\ (\ref{eqXCov}), the covariance ellipses for $\{ \widehat{\bSigma}_{s} \}$ contain $\widehat{\bSigma}$ in their intersection, for any values of the cross-covariance terms. Hence the method yields
\begin{align}
\widehat{\bSigma}^{-1} &= \sum_{s} \omega_{s} \widehat{\bSigma}_{s}^{-1}, \label{fuseCI1} \\ 
\widehat{\bmu} &= \widehat{\bSigma} \sum_{s} \omega_{s} \widehat{\bSigma}_{s}^{-1} \widehat{\bmu}_{s} , \label{fuseCI2}
\end{align}
for scalars $\omega_s \in [0,1]$ with $\sum_s \omega_s = 1$, since a convex combination of covariances ensures $\widehat{\bSigma}$ encloses the intersection region.  The $\omega_s$ terms can be computed via a standard optimization routine subject to minimising the trace of $\widehat{\bSigma}$.  Further, it has been shown that the CI method applied to a pair of consistent estimates yields a consistent estimate \cite{Uhl95}.  Finally, equations (\ref{fuseCI1}) and (\ref{fuseCI2}) as given above extend the canonical case of two sources, as offered in \cite{Chen02}.

\vspace{0.25 \baselineskip}

\noindent CI is a straightforward method that holds for any cross-covariance, but a cost of its flexibility is that the estimate can be too conservative for many practical use cases.  We next consider Inverse Covariance Intersection that attempts to retain the consistency benefit of CI, while being less conservative.

\subsubsection{Inverse Covariance Intersection}

The method of inverse Covariance Intersection (ICI) is similar in approach to that of CI, but achieves a less conservative estimate while still maintaining consistency.  The algorithm was presented recently in \cite{Noa17}.  A central insight, inspired by \cite{Sij12}, is to first consider estimates pairwise, such that for $s=1,2$:
\begin{align}
\widehat{\bmu}_{s} &= \widehat{\bSigma}_{s} \Big( (\widetilde{\bSigma}_{s})^{-1} \widetilde{\bmu}_{s} + \bGamma^{-1} \bgamma \Big), \label{fuseDecomp1}\\
\widehat{\bSigma}_{s} &= \Big((\widetilde{\bSigma}_{s})^{-1} + \bGamma^{-1}\Big)^{-1}. \label{fuseDecomp2}
\end{align}
This is to say, that each estimate is itself the PW fusion of a sub-estimate from the collection $\{ (\widetilde{\bmu}_s, \widetilde{\bSigma}_s): 1 \le s \le 2 \}$, for sub-estimates uncorrelated between sources, and a shared set of information denoted  $(\bgamma,\bGamma)$, also uncorrelated with each sub-estimate;  we proceed assuming such a decomposition exist.  If the shared information is known, then fusion can be solved optimally by subtracting the shared covariance from the PW fusion covariance for the original pair of estimates \cite{Noa17}.  But since it is not known, the second insight is to instead solve for, and subtract, an upper bound for the shared covariance, valid for any true underlying shared information.  Further, regardless of the shared information, it is shown that a consistent, tight covariance estimate -- if it exists -- is of the form
\begin{align}
\widehat{\bSigma}^{-1} = \sum_{i=1}^2 \widehat{\bSigma}_{s}^{-1} - \Big( \sum_{s=1}^2 \omega_{s} \widehat{\bSigma}_{s} \Big)^{-1},  \label{fuseICI1}
\end{align}
for scalars $\omega_s$ as in the case of CI, with the ICI covariance estimate achieving this form exactly  \cite{Noa17}.  With a modicum of algebra, it follows that the fused mean is given by
\begin{align}
\widehat{\bmu} &= \sum_{s=1}^2 C_{s} \widehat{\bmu}_{s},  \label{fuseICI2}\\
\text{where } C_{s} &= \widehat{\bSigma} \cdot \Big( \widehat{\bSigma}_{s}^{-1} - \omega_s \Big( \sum_{s=1}^2 \omega_s \widehat{\bSigma}_{s}  \Big)^{-1}\Big).  \nonumber
\end{align}

\noindent It is important to regonise that this formula does not generalise to more than two information sources in an obvious way -- the ICI algorithm fuses information sources pairwise.  However, ICI can be applied recursively. This follows from the assumption that any pairwise-fused estimate has a decomposition analogous to equations (\ref{fuseDecomp1}) and (\ref{fuseDecomp2}) \cite{Noa17}.  For further results and details we also recommend \cite{Noa17_2}.

\subsubsection{Covariance Union}

The fusion methods outlined above assume that each source's estimate is itself consistent with respect to the target.  We relax this assumption by requiring that at least one of the source estimates is consistent.  A given estimate could be inconsistent, as evidenced by, say, mean components sufficiently different between estimates, given  relatively smaller covariance estimates.  This could be apparent, for example, when the Mahalanobis distance between some pair of estimates exceeds a critical threshold.  Further, we may not know precisely which of the underlying estimates is consistent, or it may be (in some sense) costly to resolve.  The Covariance Union (CU)  algorithm has been proposed as a fusion method for this setting \cite{Uhl03}.  CU constructs a fused estimate guaranteed to be consistent, since the fused estimate is consistent with respect to each underlying estimate:
\begin{align*}
\widehat{\bSigma} &\ge \widehat{\bSigma}_s + (\widehat{\bmu}-\widehat{\bmu}_s) (\widehat{\bmu}-\widehat{\bmu}_s)^T,  \qquad 1 \le s \le S.
\end{align*}

\newpage

\begin{center}
\makebox[\textwidth]{
  \includegraphics[height=4.2in, width=5.6in]{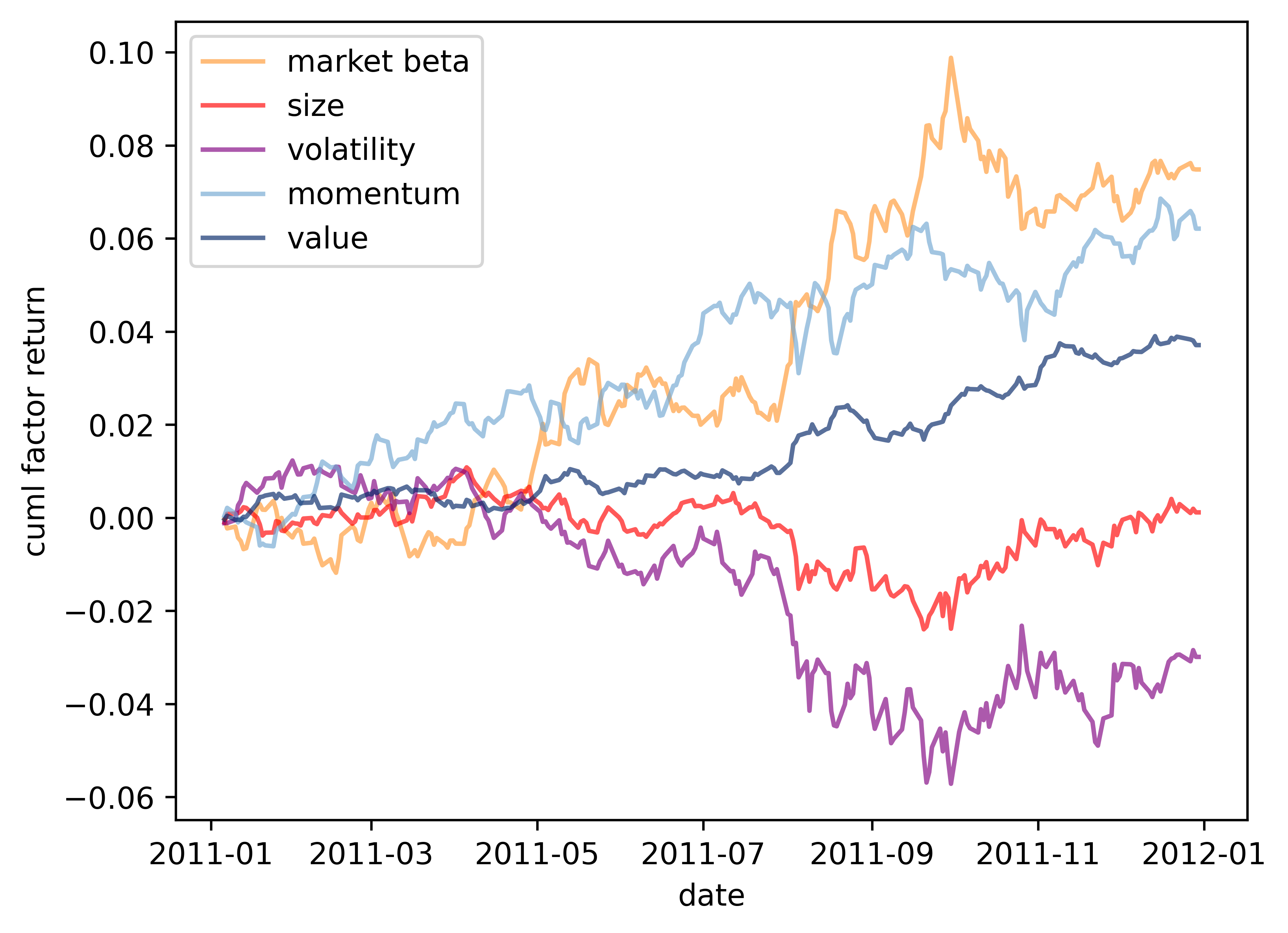}}\par
  \captionof{figure}{Cumulative factor returns to risk premia as described in Section 2.2; replication of Figure 1 of \cite{Kolm17}.  A single year is shown for ease of visualisation.  The year 2011 is arbitrarily chosen.
 \label{fg:A}}
\end{center}

\noindent While the idea behind CU is relatively simple, the numerical implementation for estimating the fused estimate is more challenging.  In an attempt to simplify the computation, we offer the following details.  Firstly, CU can be implemented using non-convex optimisation software such as SolvOpt \cite{Kun97}.  We follow \cite{Boc06} for implementing a SolvOpt-based solution for calculating the results of Section 5 below.  Secondly, a key step is to define the coefficient vector that we wish to solve for, $\bx$.  The first $n$ elements of $\bx$ are the elements of $\widehat{\bmu}$, and the remaining $\frac{n(n+1)}{2}$ elements of $\bx$ are the elements of the upper triangle of $\widehat{\bSigma}$.  Then $\bx$ becomes the optimization target for SolvOpt.  We seek to minimise the determinant of $\widehat{\bSigma}$ subject to the constraint that 
\begin{align}
\forall s, \qquad \widehat{\bSigma}_s^* := \widehat{\bSigma} &- \widehat{\bSigma}_s - (\widehat{\bmu}-\widehat{\bmu}_s) (\widehat{\bmu}-\widehat{\bmu}_s)^T \ge 0. \label{CUconstraint}
\end{align}
A particular feature of the SolvOpt software is that it requires a constraint $c(\bx)$ such that, for any input, $c(\bx) \le 0$.  Since $\widehat{\bSigma}_s^*$ is required to be positive semi-definite, it must have non-negative eigenvalues for all choices of $s$, and the constraint is satisfied for a choice of $c$ such that
$$
-c(\bx) := \min\{ \{ \text{eigenvalues}(\widehat{\bSigma}_s^*) : \forall s  \}  \} \ge 0.
$$
Finally, for the interested reader, the works of \cite{Boc06, Jul04} offer further efficient approximation methods for implementing CU, as required for particular practical use cases.

\section{Data}

\noindent  We describe the data set construction for replicating the empirical example for BL-APT of \cite{Kolm17}.  This construction matches the original descriptions as closely as possible, and we make note of any additional required assumptions made.  While we have attempted to carefully reproduce the database, our results need not match exactly.  Not in the least, we have no guarantee that the underlying database that we call is the same 5 years later, even if our data replication process is perfectly faithful.  A visual comparison can be made between our factor values against the source given Figure \ref{fg:A} above, which demonstrates a reasonable replication.  

\vspace{0.25 \baselineskip}

\noindent  The data is sourced fom CRSP and IBES databases with access granted via the Wharton Research Data Service.  The original work sourced daily US equity market data from 1992-2015; we extend our data set till 2022 to include the market crash of March 2020, and subsequent market rebound, for interest's sake.

\vspace{0.25 \baselineskip}

\noindent   For each day, set $n=2000$ and select the top $n$ stocks within the US market sorted by market capitalisation.  The data set is filtered for USD denominated common stock only -- there are no closed end funds, REITs, ETFs, unit trusts, depository receipts, warrants etc.  We also collected the S\&P500 (excess) return time-series, stock market capitalisations, and earnings per share, adjusted for splits.

\vspace{0.25 \baselineskip}

\noindent  The data set requires the creation of 5 risk factors: market beta, size, volatility, momentum and value.  The reasonableness of our data set construction is per the output of Figure \ref{fg:A}, which accords with expectations.  We also create 70 industry factors; these are one-hot encoded based on the industry given by SIC.  Calculating size is relatively straightforward in CRSP as the product of shares outstanding and the daily close price.  Value was determined by analysing IBES data, and could be merged back to the CRSP database based on the  CUSIP identifier.  We note that risk factor construction is often non-trivial.  Momentum was codable using CRSP data and the definitions of \cite{Asn13}; we calculated the daily compounded 12 month returns sans the most recent 1 month of data.  We created the market Beta and volatility factors using the relevant CRSP data and the details of \cite{Kolm17}.  Indeed, over a two year rolling window we estimate a collection of linear functions by regressing each asset's daily excess returns against the S\&P500 excess returns.  The market Beta is the regression gradient, while the volatility factor is set to be the regression mean-square error.  

\noindent  The data is constructed for bi-monthly rebalancing to facilitate the experiments of the following section.

\section{Results}

\emph{View models.}  To proxy multiple sources of view generation, we implemented various models with utility for forecasting financial time-series.  Indeed, machine learning models have proven useful for not only prediction, but also for estimating prediction uncertainty.  We implement an ARIMA, boosted regression and Gaussian Process model for generating views, and calculate model-derived uncertainty estimates.  Relevant details are included in the Appendix, since time-series prediction modelling is not the primary focus of this paper.   We denote the three view models `ARIMA', `Boost' and `GP', respectively.  We also implement the 4 fusion methods outlined in Section 3.

\vspace{0.25 \baselineskip}

\noindent  \emph{Parameter choices.} Per Figure \ref{fg:C} and the outline in Section 2, the BL-APT model requires the specification of $\bq, \bF,\bOmega, \bxi, \bV$ and $\bD$.  We set $\bq$ as the mean prediction estimate given by a view model, with the view models as previously described.  The accompanying epistemic uncertainty estimate is set to $\bOmega$, and $\bF$ is set equal to the aleatoric uncertainty estimate.  The prior hyperparameter $\bxi$ is estimated as an average of $\bef$ over a recent rolling lookback window of the most recent 20 observations.  This window size was arbitraily decided, and is the same for all rolling windows introduced hereafter.  The total variance of each underlying factor is calculated relative to out-of-sample data, and we yield an approximation to epistemic uncertainty by the estimation method as described for the ARIMA model in the Appendix below.  We set $\bV$ as this estimate.  Our estimate for $\bD$ is a diagonal covariance matrix equal to $\widehat{\sigma}^2 \bI$, where $\widehat{\sigma}^2$ is the mean square error yielded from estimating eq.\ (\ref{eqR}), and $\bI$ is the identity matrix.  Finally, we set the relative risk aversion parameter to a constant value of $10$.

\vspace{0.25 \baselineskip}

\noindent  \emph{Transaction costs.}  An effect of accounting for transaction costs, and carefully setting the transaction cost model parameters, is to greatly reduce portfolio turnover relative to the unconstrained Markowitz approach

\begin{center}
\makebox[\textwidth]{
  \includegraphics[width=7.9in]{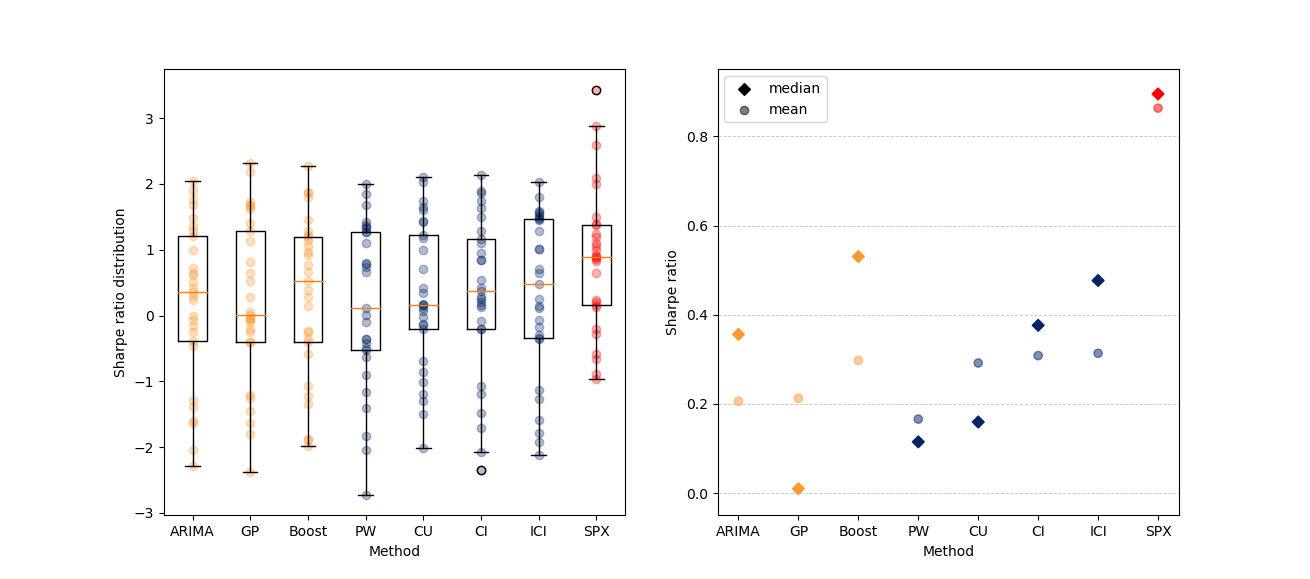}}\par
  \captionof{figure}{View-based portfolio excess return performance net of trading costs. Benchmark SPX strategy is gross of (negligible) transaction costs, but total returns are calculated in excess of the risk-free rate.
 \label{fg:TC2}}
\end{center}

\noindent with costs assumed nil.  Indeed, without transaction costs, it is well-known that this latter approach usually result in unrealistic and unreasonably large leverage suggested for the optimal portfolio, and high portfolio turnover.  Recent results of \cite{Jen22} show that incorporating an appropriate transaction cost model goes some way to mitigate these limitations in empirical studies.  We implement the model outlined in Section 2.4 above to the same ends.  We set our portfolio rebalance period to be bi-monthly, which is sufficiently long -- given our assumed wealth process, and the potential return being signficantly larger than the cost of rebalancing -- such that the reward of updating weights is not dominated by the cost of trading.

\vspace{0.25 \baselineskip}

\noindent \emph{Results -- benchmark.}  We have tabulated performance results by year and model over a 29 year period, and include them in the Appendix. The method `S\&P500 (TR)' denotes the total return of the S\&P 500 for the year.  The annual returns may appear slightly less than commonly reported since they are net of the risk-free rate.  The S\&P500 (TR) represents a useful benchmark in that it represents the market buy-and-hold strategy, accessible to the modern day investor and with a potential for exceptionally competitive fees, especially in the more recent years.  For comparisons sake, all results dependent on an underlying view model are reported based on normalised underlying data, such that ex-post annualised return volatility is equal to that of the benchmark portfolio.   We make a note that, on perusing the tables, there is obvious decay in the performance of the view-based models over the 30 year period, as evidenced, for example, by the significant decline in Sharpe ratios over time.

\vspace{0.25 \baselineskip}

\noindent \emph{Results -- statistics.}  We depict the Sharpe ratio statistics from the tabulated data in Figure \ref{fg:TC2} above.  The left-most chart shows box \& whisker plots for the annual Sharpe ratio for the collection of portfolio's over the 29 year period.  Investment in the S\&P500 (TR) is clearly superior to our underlying modeling approach.  On the other hand, there are improvements that could increase the view model performance that we deem beyond the scope of this work.  For an easier comparison of centrality for the various method's Sharpe distributions, we show each portfolio's median and mean Sharpe ratio over the 29 year period in the right-most chart.  This visual depiction summarises how fusion and non-fusion based methods performed relative to each other through time (and relative to the market total return portfolio).  We observe that fusion-based medians are increasing through PW (0.11), CU (0.16), CI (0.38) and ICI (0.48), with the means showing a similar but less pronounced trend.  The set of mean Sharpe ratios for the single-view based methods have a smaller range than the medians (0.21-0.30 versus 0.01--0.53, respectively); similarly for the fusion-based methods (0.17--0.31 versus 0.11--0.48).  When considering the mean Sharpe ratio over the full data sample, ICI outperforms amongst the set of fusion-based methods and the single-view-based models.  Similarly to the median, ICI outperforms amongst the set of fusion-based methods and sits closely to the best of the single-view-based models.   We test for statistical significance of differences in these measures of distribution centrality per the results in Figure \ref{fg:R1} below.  Despite the small sample sizes there is some interesting supporting evidence for the outperformance of fusion-based methods.

\vspace{0.25 \baselineskip}

\noindent We test for the difference of means between methods with a collection of pairwise 1-sided t-tests.  Such tests make a critical assumption of independence in the pairwise Sharpe ratio differences as indexed by year.  We note that despite individual time-series of model performance statistics potentially displaying serial correlation, it seems reasonable to suggest that the the performance between models over the years does not.  This is clearly a critical claim.  We examine the autocorrelation function for the 28 possible pairwise comparisons and find 1 case of rejecting the null hypothesis of no autocorrelation in a Ljung-Box test at the 10\% level of significance.  We further assume normality of the differences and test this with a (one-sided) Shapiro-Wilk test for normality, also at the 10\% level of significance.  We find that this assumption is rejected for 3 cases.  Finally, we assume the non-existence of outliers in the difference data, and validate this claim on visual inspection of box and whisker plots.  

\vspace{0.25 \baselineskip}

\noindent On testing for the difference of means with a paired-t test, we find, of particular interest, that ICI outperforms the ARIMA and GP models, while CI outperforms the ARIMA only (though in this case the Shaprio-Wilks test leads us not to rely on some of the t-test p-values).  To support testing for significance in the difference of means between groups, we bootstrap 2-sided bias-corrected and accelerated (BCa) bootstrap 90\% confidence intervals for paired data \cite{Efr87}.  We find evidence that ICI outperforms the ARIMA and GP models, and again the CI outperforms the ARIMA model.  We also test for the difference in medians between groups with BCa confidence intervals, but they are broadly less conclusive.  This is potentially owing to the small sample size.  On the other hand, we have evidence that both ICI and CI outperforms GP on this metric.

\vspace{0.25 \baselineskip}

\noindent Alternatively to testing for the difference of medians across all samples, we can test for the median of the differences between pairwise annual Sharpes.  To this end, we utilise the Wilcoxon signed-rank test for paired samples \cite{Wil45}.  In addition to assuming independence, a critical assumption is that of symmetry about the distributions of paired differences.  Given the results of the Shapiro-Wilk tests, and visual inspection of what appear to be reasonable QQ plots, we continue assuming symmetry is satisfied.  We perform 1-sided tests at the 10\% level of signficance and find evidence that ICI outperforms the ARIMA model for this comparison.

\vspace{0.25 \baselineskip}

\noindent  It is interesting that the CI and ICI methods achieve more compelling outperformance than the CU and PW fusion methods, relative to the underlying view models.  The assumption of PW that the underlying views have $0$ cross-covariance may be suboptimal, such that the method is yielding inconsistent estimates and in turn impacting trading performance.  On the other hand, perhaps CU could be utilised more sparingly and strategically, such as when the underlying views have relatively high dispersion and so appear sufficiently contradictory.  An alternative fusion method such as ICI could be employed otherwise.  

\vspace{0.25 \baselineskip}

\noindent  Corroborating our intuition about the relative strength of the S\&P500 (TR) strategy relative to all others, we see that SPX improves on all models -- both single-view- and fusion-based -- for almost all tests.  Sufficiently strong view models are required to compete with the market portfolio, but beyond the scope of this work.  It is perhaps surprising that view- and fusion-based models net of transaction costs have net positive Sharpe ratios at all. Indeed, we estimate global Sharpe ratios between 0.10 (PW) and 0.34 (ICI) for the entirety of the 29 year period.

\begin{landscape}
\begin{table}[ht]
\small
  \centering
\begin{tabular}{*{8}c}
  \cmidrule(lr){1-2}   \cmidrule(lr){3-4} \cmidrule(lr){5-8}
  \multicolumn{2}{c}{ \makecell{ Pairwise \\ comparison } } & Ljung-Box& Shapiro-Wilk & \makecell{ paired-t p-vals \\ (diff of means) } & \makecell{ BCa CIs \\ (diff of means) } & \makecell{ BCa CIs \\ (diff of med.) } &  \makecell{ Wilcoxon p-vals \\ (med. of diffs) }   \\
  \cmidrule(lr){1-2}   \cmidrule(lr){3-4} \cmidrule(lr){5-8}
SPX & ICI & 0.14 & 0.21 & \cellcolor{yellow!18} \bf{0.06} & \cellcolor{yellow!18} \bf{[ 0.15, 0.98 ]} & [ -0.13, 0.90 ] & \cellcolor{yellow!18} \bf{0.07} \\
  & CI & 0.14 & 0.45 & \cellcolor{yellow!18} \bf{0.05} & \cellcolor{yellow!18} \bf{[ 0.15, 0.99 ]} & [ -0.07, 0.70 ] & \cellcolor{yellow!18} \bf{0.07} \\
  & CU & 0.12 & 0.75 & \cellcolor{yellow!18} \bf{0.05} & \cellcolor{yellow!18} \bf{[ 0.19, 1.01 ]} & \cellcolor{yellow!18} \bf{[ 0.31, 1.03 ]} & \cellcolor{yellow!18} \bf{0.06} \\
  & PW & 0.17 & 0.36 & \cellcolor{yellow!18} \bf{0.03} & \cellcolor{yellow!18} \bf{[ 0.29, 1.22 ]} & \cellcolor{yellow!18} \bf{[ 0.07, 1.30 ]} & \cellcolor{yellow!18} \bf{0.06} \\
  & ARIMA & 0.13 & 0.41 & \cellcolor{yellow!18} \bf{0.03} & \cellcolor{yellow!18} \bf{[ 0.25, 1.09 ]} & \cellcolor{yellow!18} \bf{[ 0.17, 0.92 ]} & \cellcolor{yellow!18} \bf{0.03} \\
  & GP & 0.17 & 0.51 & \cellcolor{yellow!18} \bf{0.03} & \cellcolor{yellow!18} \bf{[ 0.20, 1.09 ]} & \cellcolor{yellow!18} \bf{[ 0.34, 1.10 ]} & \cellcolor{yellow!18} \bf{0.03} \\
  & Boost & 0.16 & 0.32 & \cellcolor{yellow!18} \bf{0.05} & \cellcolor{yellow!18} \bf{[ 0.18, 1.01 ]} & [ -0.10, 0.83 ] & \cellcolor{yellow!18} \bf{0.08} \\
  \cmidrule(lr){1 - 2}   \cmidrule(lr){3 - 4} \cmidrule(lr){5 - 8} 
ICI & CI & 0.80 & \cellcolor{yellow!18} \bf{0.03} & 0.47 & [ -0.08, 0.08 ] & [ -0.26, 0.28 ] & 0.31 \\
  & CU & 0.88 & 0.12 & 0.41 & [ -0.11, 0.14 ] & \cellcolor{yellow!18} \bf{[ 0.01, 0.67 ]} & 0.63 \\
  & PW & 0.94 & 0.90 & 0.15 & [ -0.03, 0.32 ] & \cellcolor{yellow!18} \bf{[ 0.04, 1.00 ]} & 0.19 \\
  & ARIMA & 0.70 & 0.75 & \cellcolor{yellow!18} \bf{0.08} & \cellcolor{yellow!18} \bf{[ 0.02, 0.21 ]} & [ -0.26, 0.36 ] & \cellcolor{yellow!18} \bf{0.10} \\
  & GP & 0.33 & 0.65 & \cellcolor{yellow!18} \bf{0.09} & \cellcolor{yellow!18} \bf{[ 0.01, 0.19 ]} & \cellcolor{yellow!18} \bf{[ 0.25, 0.73 ]} & 0.12 \\
  & Boost & 0.68 & 0.19 & 0.46 & [ -0.21, 0.25 ] & [ -0.51, 0.39 ] & 0.58 \\
  \cmidrule(lr){1 - 2}   \cmidrule(lr){3 - 4} \cmidrule(lr){5 - 8} 
CI & CU & 0.56 & 0.85 & 0.43 & [ -0.12, 0.14 ] & \cellcolor{yellow!18} \bf{[ 0.05, 0.51 ]} & 0.55 \\
  & PW & 0.67 & 0.56 & 0.15 & [ -0.02, 0.32 ] & [ -0.29, 0.65 ] & 0.18 \\
  & ARIMA & 0.47 & \cellcolor{yellow!18} \bf{0.00} & \cellcolor{yellow!18} \bf{0.06} & \cellcolor{yellow!18} \bf{[ 0.03, 0.19 ]} & [ -0.17, 0.22 ] & 0.93 \\
  & GP & 0.11 & 0.74 & 0.13 & [ -0.01, 0.20 ] & \cellcolor{yellow!18} \bf{[ 0.30, 0.67 ]} & 0.16 \\
  & Boost & 0.35 & 0.91 & 0.47 & [ -0.17, 0.24 ] & [ -0.52, 0.10 ] & 0.50 \\
  \cmidrule(lr){1 - 2}   \cmidrule(lr){3 - 4} \cmidrule(lr){5 - 8} 
CU & PW & 0.29 & 0.28 & 0.17 & [ -0.04, 0.28 ] & [ -0.57, 0.43 ] & 0.17 \\
  & ARIMA & 0.48 & 0.89 & 0.23 & [ -0.05, 0.22 ] & \cellcolor{yellow!18} \bf{[ -0.46, -0.02 ]} & 0.20 \\
  & GP & 0.75 & 0.22 & 0.23 & [ -0.06, 0.20 ] & [ -0.13, 0.37 ] & 0.21 \\
  & Boost & 0.73 & 0.98 & 0.49 & [ -0.21, 0.19 ] & [ -0.78, 0.01 ] & 0.52 \\
  \cmidrule(lr){1 - 2}   \cmidrule(lr){3 - 4} \cmidrule(lr){5 - 8} 
PW & ARIMA & 0.95 & 0.64 & 0.36 & [ -0.17, 0.11 ] & [ -0.65, 0.24 ] & 0.65 \\
  & GP & 0.95 & 0.58 & 0.33 & [ -0.19, 0.08 ] & [ -0.31, 0.66 ] & 0.41 \\
  & Boost & 0.17 & 0.45 & \cellcolor{yellow!18} \bf{0.09} & \cellcolor{yellow!18} \bf{[ -0.27, -0.03 ]} & \cellcolor{yellow!18} \bf{[ -1.12, -0.14 ]} & \cellcolor{yellow!18} \bf{0.10} \\
    \cmidrule(lr){1 - 2}   \cmidrule(lr){3 - 4} \cmidrule(lr){5 - 8} 
ARIMA & GP & \cellcolor{yellow!18} \bf{0.01} & 0.94 & 0.46 & [ -0.08, 0.07 ] & \cellcolor{yellow!18} \bf{[ 0.22, 0.54 ]} & 0.45 \\
  & Boost & 0.20 & \cellcolor{yellow!18} \bf{0.06} & 0.27 & [ -0.29, 0.09 ] & [ -0.51, 0.19 ] & 0.73 \\
  \cmidrule(lr){1 - 2}   \cmidrule(lr){3 - 4} \cmidrule(lr){5 - 8} 
GP & Boost & 0.74 & 0.34 & 0.27 & [ -0.26, 0.07 ] & \cellcolor{yellow!18} \bf{[ -0.96, -0.32 ]} & 0.64 \\
\cmidrule(lr){1 - 2}   \cmidrule(lr){3 - 4} \cmidrule(lr){5 - 8} 
  \end{tabular}
  
\captionof{figure}{Tabulated p-values and confidence intervals for a battery of statistical tests for comparing annual 29 years of  annual Sharpe ratios pairwise between strategies.  Yellow highlighting is applied for significance at the 10\% level.  The rightmost four columns represent one-sided tests.}
\label{fg:R1}
\end{table}
\end{landscape}

\vspace{0.25 \baselineskip}

\noindent \emph{Closing remark.}  The results presented offer preliminary evidence supporting fusion-based methods within a Black-Litterman framework subsuming a relatively complex investment strategy net of transaction costs.  Fusion-based methods result in a weighted-average performance respectful of uncertainty estimates.  These methods could certainly be ex-ante preferred when all views are to be incorporated into an investment decision, in the case that the best view model is unknown (and/or as in our presented work, when the method constituting the best view model is both largely unpredictable, and changing throught time).  We have supporting evidence for a rare `free lunch' in financial investing by way of model fusion, essentially amounting to a benefit we could call `model diversification'.  This could be of further application in alternative trading domains, perhaps with more sophisticated underlying view models, and we hope to inspire readers to consider their practical use.

\section{Conclusion and next steps}

In this work, we further characterise view uncertainty within a Bayesian Black-Litterman framework.  In doing so, we offer optimal portfolio weights for a case of Black-Litterman in an Arbitrage Pricing Theory setting.  Considering the Bayesian Black-Litterman model via an equivalent state-space representation, we are inspired by the information fusion literature for combining correlated prediction mean and uncertainty estimates streamed from diverse sources.  Hence, we demonstrate prediction and uncertainty fusion for multiple, simultaneous views within the Black-Litterman framework, and describe consistent uncertainty estimation within financial investing contexts.  In our empirical work, we demonstrate the utility of our approach for BL-APT.  Future practical work could be based on more realistic investment strategies, not the least incorporating methods of drawdown control.

\section*{Acknowledgements}

The authors would like to thank the Oxford-Man Institute of Quantitative Finance for its generous support.  Professor Stephen Roberts would like to further thank the Royal Academy of Engineering.  Trent Spears would like to further thank Professor Nir Vulkan and Dr Jan-Peter Calliess for their helpful comments and suggestions.

\bibliographystyle{unsrtnat}
\setlength{\bibsep}{1pt}
\footnotesize\bibliography{bibs}{}

\begin{thebibliography}{42}
\providecommand{\natexlab}[1]{#1}
\providecommand{\url}[1]{\texttt{#1}}
\expandafter\ifx\csname urlstyle\endcsname\relax
  \providecommand{\doi}[1]{doi: #1}\else
  \providecommand{\doi}{doi: \begingroup \urlstyle{rm}\Url}\fi

\bibitem[Black and Litterman(1991)]{Black91}
Fischer Black and Robert~B. Litterman.
\newblock Asset {A}llocation: {C}ombining {I}nvestor {V}iews with {M}arket
  {E}quilibrium.
\newblock \emph{The Journal of Fixed Income}, 1\penalty0 (2):\penalty0 7--18,
  1991.

\bibitem[Walters(2014)]{Jay14}
Jay Walters.
\newblock The black-litterman model in detail.
\newblock \url{https://ssrn.com/abstract=1314585}, 2014.
\newblock [Online; last accessed 10-Aug-2022].

\bibitem[Satchell and Scowcroft(2000)]{Sat00}
Stephen Satchell and Alan Scowcroft.
\newblock A demystification of the black–litterman model: Managing
  quantitative and traditional portfolio construction.
\newblock \emph{Journal of Asset Management}, 1:\penalty0 138--150, 2000.

\bibitem[Kolm and Ritter(2017)]{Kolm17}
Petter~N. Kolm and Gordon Ritter.
\newblock On the {B}ayesian interpretation of {B}lack–{L}itterman.
\newblock \emph{European Journal of Operational Research}, 258\penalty0
  (2):\penalty0 564--572, 2017.

\bibitem[Sharpe(1964)]{Sharpe64}
William~F. Sharpe.
\newblock Capital asset prices: A theory of market equilibrium under conditions
  of risk.
\newblock \emph{The Journal of Finance}, 19\penalty0 (3):\penalty0 425--442,
  1964.

\bibitem[Martin van~der Schans(2017)]{Sch17}
Hens~Steehouwer Martin van~der Schans.
\newblock Time-dependent black–litterman.
\newblock \emph{Journal of Asset Management}, 18:\penalty0 371--387, 2017.

\bibitem[Giglio et~al.(2022)Giglio, Kelly, and Xiu]{Gig22}
Stefano Giglio, Bryan Kelly, and Dacheng Xiu.
\newblock Factor models, machine learning, and asset pricing.
\newblock \emph{Annual Review of Financial Economics}, 14\penalty0
  (1):\penalty0 337--368, 2022.

\bibitem[Ross(1976)]{Ross76}
Stephen~A. Ross.
\newblock The arbitrage theory of capital asset pricing.
\newblock \emph{Journal of Economic Theory}, 13\penalty0 (3):\penalty0
  341--360, 1976.

\bibitem[Markowitz(1952)]{Ma52}
Harry Markowitz.
\newblock Portfolio {S}election.
\newblock \emph{The Journal of Finance}, 7\penalty0 (1):\penalty0 77--91, 1952.

\bibitem[Fabozzi et~al.(2006)Fabozzi, Focardi, and Kolm]{Fab06}
Frank~J. Fabozzi, Sergio~M. Focardi, and Petter~N. Kolm.
\newblock Incorporating {T}rading {S}trategies in the {B}lack-{L}itterman
  {F}ramework.
\newblock \emph{The Journal of Trading}, 1\penalty0 (2):\penalty0 28--37, 2006.

\bibitem[Murphy(2012)]{Murphy12}
Kevin~P. Murphy.
\newblock \emph{Machine Learning: A Probabilistic Perspective}.
\newblock The MIT Press, 2012.

\bibitem[Jensen et~al.(2022)Jensen, Kelly, Malamud, and Pedersen]{Jen22}
Theis~I. Jensen, Bryan~T. Kelly, Semyon Malamud, and Lasse~H. Pedersen.
\newblock Machine learning and the implementable efficient frontier, swiss
  finance institute research paper no. 22-63.
\newblock \url{https://ssrn.com/abstract=4187217}, 2022.
\newblock [Online; last accessed 10-Aug-2022].

\bibitem[G\^{a}rleanu and Pedersen(2013)]{Gar13}
Nicolae G\^{a}rleanu and Lasse~H. Pedersen.
\newblock Dynamic trading with predictable returns and transaction costs.
\newblock \emph{The Journal of Finance}, 68\penalty0 (6):\penalty0 2309--2340,
  2013.

\bibitem[Frazzini et~al.(2018)Frazzini, Israel, and Moskowitz]{Fraz18}
Andrea Frazzini, Ronen Israel, and Tobias~J. Moskowitz.
\newblock Trading costs.
\newblock \url{https://ssrn.com/abstract=3229719}, 2018.
\newblock [Online; last accessed 10-Aug-2022].

\bibitem[Date and Ponomareva(2010)]{Date10}
Paresh Date and Ksenia Ponomareva.
\newblock Linear and non-linear filtering in mathematical finance: {A} review.
\newblock \emph{IMA Journal of Management Mathematics}, 21, 06 2010.

\bibitem[Willner et~al.(1976)Willner, Chang, and Dunn]{Wil76}
Dieter Willner, Chaw-Bing Chang, and Keh-Ping Dunn.
\newblock Kalman filter algorithms for a multi-sensor system.
\newblock \emph{1976 IEEE Conference on Decision and Control including the 15th
  Symposium on Adaptive Processes}, pages 570--574, 1976.

\bibitem[Khaleghi et~al.(2013)Khaleghi, Khamis, Karray, and Razavi]{Kha11}
Bahador Khaleghi, Alaa Khamis, Fakhreddine~O. Karray, and Saiedeh~N. Razavi.
\newblock Multisensor data fusion: {A} review of the state-of-the-art.
\newblock \emph{Information Fusion}, 14\penalty0 (1):\penalty0 28--44, 2013.

\bibitem[Castanedo(2013)]{Cas13}
Federico Castanedo.
\newblock A review of data fusion techniques.
\newblock \emph{The Scientific World Journal}, 2013.

\bibitem[Uhlmann(1995)]{Uhl95}
Jeffrey~K. Uhlmann.
\newblock \emph{Dynamic Map Building and Localization : New Theoretical
  Foundations}.
\newblock PhD thesis, University of Oxford, 1995.

\bibitem[Thorp(2006)]{Thorp06}
Edward~O. Thorp.
\newblock The {K}elly {C}riterion in {B}lackjack, {S}ports {B}etting, and the
  {S}tock {M}arket.
\newblock \emph{Handbook of Asset and Liability Management}, 1:\penalty0
  385--428, 2006.

\bibitem[Reece and Roberts(2010)]{Ree10}
Steven Reece and Stephen Roberts.
\newblock Generalised {C}ovariance {U}nion: {A} {U}nified {A}pproach to
  {H}ypothesis {M}erging in {T}racking.
\newblock \emph{IEEE Transactions on Aerospace and Electronic Systems}, 46,
  2010.

\bibitem[Dempster(1969)]{Dem69}
Arthur~P. Dempster.
\newblock \emph{Elements of continuous multivariate analysis}.
\newblock Addison-Wesley Pub. Co., 1969.

\bibitem[Maybeck(1979)]{May79}
Peter~S. Maybeck.
\newblock \emph{Stochastic models, estimation, and control, Vol. 1}.
\newblock Academic Press, New York, 1979.

\bibitem[Bar-Shalom(1981)]{YBS81}
Yaakov Bar-Shalom.
\newblock On the track-to-track correlation problem.
\newblock \emph{IEEE Transactions on Automatic Control}, 26\penalty0
  (2):\penalty0 571--572, 1981.

\bibitem[Julier and Uhlmann(1997)]{Uhl97}
Simon~J. Julier and Jeffrey~K. Uhlmann.
\newblock A non-divergent estimation algorithm in the presence of unknown
  correlations.
\newblock In \emph{Proceedings of the 1997 American Control Conference},
  volume~4, pages 2369--2373. IEEE, 1997.

\bibitem[Chen et~al.(2002)Chen, Arambel, and Mehra]{Chen02}
Lingji Chen, P.O. Arambel, and R.K. Mehra.
\newblock Fusion under unknown correlation - covariance intersection as a
  special case.
\newblock In \emph{Proceedings of the Fifth International Conference on
  Information Fusion. FUSION 2002.}, volume~2, pages 905--912, 2002.

\bibitem[Noack et~al.(2017{\natexlab{a}})Noack, Sijs, Reinhardt, and
  Hanebeck]{Noa17}
Benjamin Noack, Joris Sijs, Marc Reinhardt, and Uwe~D. Hanebeck.
\newblock Decentralized data fusion with inverse covariance intersection.
\newblock \emph{Automatica}, 79:\penalty0 35--41, 2017{\natexlab{a}}.

\bibitem[Sijs and Lazar(2012)]{Sij12}
Joris Sijs and Mircea Lazar.
\newblock State fusion with unknown correlation: Ellipsoidal intersection.
\newblock \emph{Automatica}, 48\penalty0 (8):\penalty0 1874--1878, 2012.

\bibitem[Noack et~al.(2017{\natexlab{b}})Noack, Sijs, and Hanebeck]{Noa17_2}
Benjamin Noack, Joris Sijs, and Uwe Hanebeck.
\newblock Inverse covariance intersection: New insights and properties.
\newblock In \emph{Proceedings of the 20th International Conference on
  Information Fusion}, 2017{\natexlab{b}}.

\bibitem[Uhlmann(2003)]{Uhl03}
Jeffrey~K. Uhlmann.
\newblock Covariance consistency methods for fault-tolerant distributed data
  fusion.
\newblock \emph{Information Fusion}, 4:\penalty0 201--215, 2003.

\bibitem[Kuntsevich and Kappel(1997)]{Kun97}
Alexei Kuntsevich and Franz Kappel.
\newblock Solvopt: The solver for local nonlinear optimization problems.
\newblock \emph{Institute for Mathematics, University of Graz}, 1997.

\bibitem[Bochardt et~al.(2006)Bochardt, Calhoun, Uhlmann, and Julier]{Boc06}
Ottmar Bochardt, Ryan Calhoun, Jeffrey~K. Uhlmann, and Simon~J. Julier.
\newblock Generalized information representation and compression using
  covariance union.
\newblock \emph{2006 9th International Conference on Information Fusion}, pages
  1--7, 2006.

\bibitem[Julier et~al.(2004)Julier, Uhlmann, and Nicholson]{Jul04}
Simon~J. Julier, Jeffrey~K. Uhlmann, and David Nicholson.
\newblock A method for dealing with assignment ambiguity.
\newblock \emph{Proceedings of the 2004 American Control Conference},
  5:\penalty0 4102--4107 vol.5, 2004.

\bibitem[Asness et~al.(2013)Asness, Moskowitz, and Pedersen]{Asn13}
Clifford~S. Asness, Tobias~J. Moskowitz, and Lasse~Heje Pedersen.
\newblock Value and {M}omentum {E}verywhere.
\newblock \emph{The Journal of Finance}, 68\penalty0 (3):\penalty0 929--985,
  2013.

\bibitem[Efron(1987)]{Efr87}
Bradley Efron.
\newblock Better bootstrap confidence intervals.
\newblock \emph{Journal of the American Statistical Association}, 82\penalty0
  (397):\penalty0 171--185, 1987.

\bibitem[Wilcoxon(1945)]{Wil45}
Frank Wilcoxon.
\newblock Individual comparisons by ranking methods.
\newblock \emph{Biometrics Bulletin}, 1\penalty0 (6):\penalty0 80--83, 1945.

\bibitem[H{\"{u}}llermeier and Waegeman(2021)]{Eyke19}
Eyke H{\"{u}}llermeier and Willem Waegeman.
\newblock Aleatoric and epistemic uncertainty in machine learning: an
  introduction to concepts and methods.
\newblock \emph{Machine Learning}, 110:\penalty0 457--506, 2021.

\bibitem[Pedregosa et~al.(2011)Pedregosa, Varoquaux, Gramfort, Michel, Thirion,
  Grisel, Blondel, Prettenhofer, Weiss, Dubourg, Vanderplas, Passos,
  Cournapeau, Brucher, Perrot, and Duchesnay]{scikit-learn}
F.~Pedregosa, G.~Varoquaux, A.~Gramfort, V.~Michel, B.~Thirion, O.~Grisel,
  M.~Blondel, P.~Prettenhofer, R.~Weiss, V.~Dubourg, J.~Vanderplas, A.~Passos,
  D.~Cournapeau, M.~Brucher, M.~Perrot, and E.~Duchesnay.
\newblock Scikit-learn: {M}achine {L}earning in {P}ython.
\newblock \emph{Journal of Machine Learning Research}, 12:\penalty0 2825--2830,
  2011.

\bibitem[Roberts et~al.(2013)Roberts, Osborne, Ebden, Reece, Gibson, and
  Aigrain]{Rob13}
Stephen Roberts, Michael Osborne, Mark Ebden, Steven Reece, N~Gibson, and
  Suzanne Aigrain.
\newblock Gaussian processes for time-series modelling.
\newblock \emph{Philosophical transactions. Series A, Mathematical, physical,
  and engineering sciences}, 371:\penalty0 20110550, 02 2013.

\bibitem[Prokhorenkova et~al.(2018)Prokhorenkova, Gusev, Vorobev, Dorogush, and
  Gulin]{Pro18}
Liudmila Prokhorenkova, Gleb Gusev, Aleksandr Vorobev, Anna~Veronika Dorogush,
  and Andrey Gulin.
\newblock Catboost: unbiased boosting with categorical features.
\newblock In \emph{Advances in Neural Information Processing Systems},
  volume~31, 2018.

\bibitem[Ustimenko et~al.(2020)Ustimenko, Ostroumova~Prokhorenkova, and
  Malinin]{Ust20}
Aleksei Ustimenko, Liudmila Ostroumova~Prokhorenkova, and Andrey Malinin.
\newblock Uncertainty in gradient boosting via ensembles.
\newblock 06 2020.

\bibitem[Lahlou et~al.(2022)Lahlou, Jain, Nekoei, Butoi, Bertin, Rector-Brooks,
  Korablyov, and Bengio]{deup22}
Salem Lahlou, Moksh Jain, Hadi Nekoei, Victor Butoi, Paul Bertin, Jarrid
  Rector-Brooks, Maksym Korablyov, and Yoshua Bengio.
\newblock Deup: Direct epistemic uncertainty prediction.
\newblock \url{https://arxiv.org/abs/2102.08501}, 2022.
\newblock [Online; last accessed 04-Dec-2022].

\end{thebibliography}

\newpage

\appendix
\small

\section{Appendix}

\subsection{Further APT calculations}

This section adds further detail to the calculations of Section 2.3.  In determining the expectation and covariance for $\br | \bq$ in the case of unknown mean but known covariance, given eq.\ (\ref{eqTMP3}) we collect and expand the terms in the exponent up to a factor of $-\frac{1}{2}$, complete the square in $\bef$, and integrate what is recognisable as a Gaussian integral:
\begin{align*}
(\br - \bX\bef)^T \bD^{-1}(\br - \bX\bef) + (\bef - \bmu)^T \bSigma^{-1} (\bef - \bmu) &= \br^T \bD^{-1} \br - 2\bef^T \bX^T\bD^{-1} \br \\
&\qquad + (\bX\bef)^T \bD^{-1} \bX\bef  + \bef^T \bSigma^{-1} \bef - 2\bef^T \bSigma^{-1} \bmu + \bmu^T \bSigma^{-1} \bmu \\
&= \br^T \bD^{-1} \br + \bef^T \bH \bef - 2 \bbeta \bef + \bmu^T \bSigma^{-1} \bmu
\end{align*}
where $\bmu$ and $\bSigma$ denote the mean and covariance for $\bef | \bq$, $\bH := \bX^T\bD^{-1}\bX + \bSigma^{-1}$ and $\bbeta := \bX^T \bD^{-1} \br +  \bSigma^{-1} \bmu$. Completing the square,
$$
\bef^T \bH \bef - 2 \bef^T \bbeta = (\bef-\bnu)^T \bH(\bef-\bnu) - \bnu^T \bH \bnu, \qquad \bnu = \bH^{-1} \bbeta,
$$
whereby
$$
\int \exp\Big( -\frac{1}{2} (\bef-\bnu)^T \bH (\bef-\bnu) \Big) d \bef = \sqrt{\frac{(2\pi)^k}{|\bH|}}.
$$
\noindent Therefore
\begin{align}
p( \br | \bq ) \propto \exp\Big( -\frac{1}{2}\Big[\br^T \bD^{-1}\br + \bmu^T \bSigma^{-1} \bmu - \bnu^T \bH \bnu \Big]\Big) = \exp\Big( -\frac{1}{2}\Big[\br^T \bD^{-1}\br + \bmu^T \bSigma^{-1} \bmu - \bbeta^T \bH^{-1} \bbeta \Big]\Big), \label{eqrgq}
\end{align}
and we retrieve the exponent of the Gaussian for $\br | \bq$.  Inspired by \cite{Kolm17}, we next make use of the following Lemma:
\begin{lemma}
If a multivariate normal random variable $\btheta$ has density $p(\btheta)$ and $-2 \log p(\btheta) = \btheta' \bH \btheta - 2 \bnu ' \btheta + (\text{terms without } \btheta)$ then $\text{var}(\btheta) = \bH^{-1}$ and $\E \btheta = \bH^{-1} \bnu$.
\end{lemma}
\noindent Collecting quadratic terms in $\br$ from the exponent of eq.\ (\ref{eqrgq}):
\begin{align}
- \bbeta^T \bH^{-1} \bbeta + \br^T \bD^{-1}\br &= - [\bX^T \bD^{-1} \br + \bSigma^{-1} \bmu]^T \bH^{-1} [\bX^T \bD^{-1} \br + \bSigma^{-1} \bmu] +  \br^T \bD^{-1}\br \nonumber \\
&= - \br^T \bD^{-1} \bX \bH^{-1} \bX^T \bD^{-1} \br + \br^T \bD^{-1} \br + \text{lower-order terms in $\br$}  \label{eqDots} \\
&= \br^T [- \bD^{-1} \bX \bH^{-1} \bX^T \bD ^{-1} + \bD^{-1}] \br + \text{lower-order terms in $\br$} \nonumber
\end{align}
Making use of the lemma, we have that the posterior predictive distribution $p (\br | \bq)$ is multivariate normal with covariance
\begin{align*}
\text{var}(\br | \bq) &= [-\bD^{-1} \bX \bH^{-1} \bX^T \bD^{-1} + \bD^{-1}]^{-1}, \\
&= [ \bD^{-1} - \bD^{-1} \bX [\bX^T \bD^{-1} \bX + [(\bV^{-1} + \bOmega^{-1})^{-1}+\bF]^{-1}]^{-1} \bX^T \bD^{-1} ]^{-1}.
\end{align*}
Futher, we collect the linear terms  in $\br$ from eq.\ (\ref{eqDots}): 
\begin{align*}
- \br^T \bD^{-1} \bX^T \bH^{-1} \bSigma^{-1} \bmu - \bmu^T \bSigma^{-1} \bH^{-1} \bX^T \bD^{-1} \br &= - 2 \bmu^T \bSigma^{-1} \bH^{-1} \bX^T \bD^{-1} \br,
\end{align*}
so that again by the lemma
\begin{align*}
\E(\br| \bq) &= \text{var}( \br | \bq ) \bD^{-1} \bX \bH^{-1} \bSigma^{-1} \bmu \\
&= \text{var}( \br | \bq ) \bD^{-1}\bX [\bX^T \bD^{-1} \bX + [(\bV^{-1} + \bOmega^{-1})^{-1}+\bF]^{-1}]^{-1} \\
& \qquad \cdot  [(\bV^{-1} + \bOmega^{-1})^{-1} + \bF]^{-1} (\bV^{-1} + \bOmega^{-1})^{-1} (\bV^{-1} \bxi + \bOmega^{-1} \bq).
\end{align*}

\newpage 

\subsection{View fusion under model-based forecasts}

Suppose we model $y = f(\bx) + \epsilon$ for some function $f$ parameterised by $\btheta$, with $\beps \thicksim N(0, \sigma^2)$.  Then we can write $y | \bx, \btheta \thicksim N(f(\bx),\sigma^2)$.  Assuming $\btheta$ has prior distribution $p(\btheta)$, then for a collection of data $\mathcal{D}$ we can update the prior via the likelihood analogous to eq.\ (\ref{eqUpdate}) above: $p(\btheta | \mathcal{D}) \propto p(\mathcal{D}|\btheta) p(\btheta)$.  Then for a new data point $\bx^*$, the predictive distribution for the corresponding $y^*$ can be written
$$
p(y^* | \bx^*, \mathcal{D}) = \int p(y^* | \bx^*, \btheta) p(\btheta | \mathcal{D}) d\btheta.
$$
Further, the variance of the prediction, $\text{var}(y^* | \bx^*, \mathcal{D})$, can be decomposed making use of the law of total variance, that is:
\begin{align}
\text{var}(y^* | \bx^*, \mathcal{D}) = \text{var}_{\btheta|\mathcal{D}}\big( \E(y^* | \bx^*, \btheta) \big) + \E_{\btheta|\mathcal{D}} \big( \text{var}(y^* | \bx^*, \btheta) \big). \label{UCdecomp}
\end{align}
Here the subscript on the outer variance and expectation terms on the right-hand side denotes that the integral is calculated with respect to the conditional density for $\btheta|\mathcal{D}$.  The first term on the right-hand side we call the \emph{epistemic} uncertainty, sometimes described as a `model' uncertainty associated with estimating the model parameters, and that is reducible with increasing data.  The second term we call the \emph{aleatoric} uncertainty; this models the underlying stochasticity of the data distribution, and is irreducible.

\vspace{0.25 \baselineskip}

\noindent Estimation of predictive distributions or the components of model uncertainty is not always tractable; consequently, various approximation methods exists for estimating these quantities.  A review is beyond the scope of this paper; we instead refer interested readers to the recent work of \cite{Eyke19} for further details.

\subsubsection{Brief overview of view models}

To yield the results of Section 5 above we first implement three diverse view models.  For completeness, we provide the implementation details below.

\vspace{0.25 \baselineskip}

\noindent We implement a Gaussian Process model using scikit-learn in Python \cite{scikit-learn}.  The kernel function is given by the sum of a radial basis function and a white-noise kernel;  see, for example, \cite{Rob13} for context on Gaussian Processes for time-series modelling.  We re-fit the model on a bi-monthly basis on a rolling data window of 1 year.  The predict method is used to generate prediction estimates and an accompanying (total) uncertainty estimate.  The uncertainty can be decomposed into epistemic and aleatoric components by setting the latter as the noise-level parameter estimated for the white-noise kernel.  By eq.\ (\ref{UCdecomp}) above, the epistemic uncertainty is then the difference between the total uncertainty and the aleatoric uncertainty estimates.

\vspace{0.25 \baselineskip}

\noindent We implement a boosted regression model making use of CatBoost \cite{Pro18}.  We re-fit the model on a bi-monthly basis on a rolling data window of 2 years.  The model yields an aleatoric uncertainty estimate owing to modelling this uncertainty directly via the loss function.  An estimate for epistemic uncertainty is approximated by the variance of the underlying ensemble of predictions.  Further details of this method can be found in \cite{Ust20}.

\vspace{0.25 \baselineskip}

\noindent We estimate an ARIMA model for view generation as a well-known baseline model.  Further, this model naturally yields an aleatoric uncertainty estimate.  To approximate epistemic uncertainty for this model, we are inspired by Algorithm 3 of \cite{deup22}.  We note that our method is somewhat ad hoc, and has subtle differences:  we estimate prediction error variance on a recent out-of-sample lookback window of size 6 months, estimating epistemic uncertainty as the difference between this term and today's aleatoric uncertainty estimate, to a minimum of 1e-8.  We re-fit the model on a bi-monthly basis on a rolling data window of 1 year.

\newpage

\subsection{Performance results by year}

Performance metrics for market total return (in excess of the risk-free rate), against a collection of Black-Litterman models with varying underlying prediction methods to generate and/or fuse views.  Methods sans the index allocation are normalised by return volatility to equal the ex-post annual return volatility of the market index.

\begin{table}[h!t]
\footnotesize
  \centering
\makebox[\textwidth]{\begin{tabular}{*{13}c}

&  \multicolumn{6}{c}{1993} & \multicolumn{6}{c}{1994}\\
  \cmidrule(lr){1-1} \cmidrule(lr){2-7} \cmidrule(lr){8-13}
  Portfolio method & Cuml Ret & Ret Vol  & Sharpe & IR & Sortino & Max DD & Cuml Ret & Ret Vol  & Sharpe & IR & Sortino & Max DD \\  
  \cmidrule(lr){1-1} \cmidrule(lr){2-7} \cmidrule(lr){8-13}
S\&P 500 (TR) & 7.11 & 8.59 & 0.82 & -1.94 & 1.21 & -8.84 & -2.03 & 9.80 & -0.21 & -1.42 & -0.28 & -8.91 \\
  \cmidrule(lr){1-1} \cmidrule(lr){2-7} \cmidrule(lr){8-13}
ARIMA & -17.50 & 8.59 & -2.04 & -2.68 & -2.01 & -16.43 & 20.00 & 9.80 & 2.04 & 1.73 & 5.65 & -18.76 \\
GP & -15.42 & 8.59 & -1.79 & -2.60 & -1.78 & -15.90 & 21.39 & 9.80 & 2.18 & 1.78 & 5.91 & -19.92 \\
Boost & -9.14 & 8.59 & -1.06 & -1.84 & -1.18 & -10.40 & 12.53 & 9.80 & 1.28 & 1.04 & 2.73 & -13.63 \\
  \cmidrule(lr){1-1} \cmidrule(lr){2-7} \cmidrule(lr){8-13}
PW & -12.04 & 8.59 & -1.40 & -2.20 & -1.44 & -12.37 & 13.90 & 9.80 & 1.42 & 1.16 & 3.07 & -14.73 \\
CU & -11.11 & 8.59 & -1.29 & -1.98 & -1.47 & -13.29 & 20.65 & 9.80 & 2.11 & 1.74 & 5.76 & -19.13 \\
CI & -20.11 & 8.59 & -2.34 & -2.81 & -2.22 & -18.60 & 18.24 & 9.80 & 1.86 & 1.59 & 6.13 & -17.71 \\
ICI & -16.51 & 8.59 & -1.92 & -2.46 & -1.82 & -16.14 & 19.94 & 9.80 & 2.03 & 1.70 & 8.08 & -18.73 \\
  \cmidrule(lr){1-1} \cmidrule(lr){2-7} \cmidrule(lr){8-13}
  \end{tabular}}


\makebox[\textwidth]{\begin{tabular}{*{13}c}

&  \multicolumn{6}{c}{1995} & \multicolumn{6}{c}{1996}\\
  \cmidrule(lr){1-1} \cmidrule(lr){2-7} \cmidrule(lr){8-13}
  Portfolio method & Cuml Ret & Ret Vol  & Sharpe & IR & Sortino & Max DD & Cuml Ret & Ret Vol  & Sharpe & IR & Sortino & Max DD \\  
  \cmidrule(lr){1-1} \cmidrule(lr){2-7} \cmidrule(lr){8-13}
S\&P 500 (TR) & 26.80 & 7.81 & 3.43 & -2.57 & 5.72 & -24.15 & 16.31 & 11.75 & 1.38 & -1.60 & 1.96 & -18.96 \\
  \cmidrule(lr){1-1} \cmidrule(lr){2-7} \cmidrule(lr){8-13}
ARIMA & -12.74 & 7.81 & -1.63 & -4.48 & -1.67 & -14.11 & 14.18 & 11.75 & 1.21 & -0.33 & 2.78 & -19.34 \\
GP & -11.30 & 7.81 & -1.45 & -4.33 & -1.52 & -13.81 & 9.49 & 11.75 & 0.81 & -0.59 & 1.65 & -16.73 \\
Boost & -15.45 & 7.81 & -1.98 & -4.56 & -2.06 & -14.93 & 10.90 & 11.75 & 0.93 & -0.49 & 1.62 & -18.25 \\
  \cmidrule(lr){1-1} \cmidrule(lr){2-7} \cmidrule(lr){8-13}
PW & -21.30 & 7.81 & -2.73 & -5.36 & -2.46 & -20.11 & 9.34 & 11.75 & 0.79 & -0.60 & 1.34 & -19.69 \\
CU & -11.64 & 7.81 & -1.49 & -4.53 & -1.57 & -14.14 & 20.40 & 11.75 & 1.74 & 0.02 & 4.69 & -22.35 \\
CI & -13.29 & 7.81 & -1.70 & -4.87 & -1.77 & -14.72 & 25.06 & 11.75 & 2.13 & 0.29 & 6.07 & -25.21 \\
ICI & -12.45 & 7.81 & -1.59 & -4.78 & -1.67 & -14.07 & 21.15 & 11.75 & 1.80 & 0.06 & 4.78 & -23.72 \\
  \cmidrule(lr){1-1} \cmidrule(lr){2-7} \cmidrule(lr){8-13}
  \end{tabular}}


\makebox[\textwidth]{\begin{tabular}{*{13}c}

&  \multicolumn{6}{c}{1997} & \multicolumn{6}{c}{1998}\\
  \cmidrule(lr){1-1} \cmidrule(lr){2-7} \cmidrule(lr){8-13}
  Portfolio method & Cuml Ret & Ret Vol  & Sharpe & IR & Sortino & Max DD & Cuml Ret & Ret Vol  & Sharpe & IR & Sortino & Max DD \\  
  \cmidrule(lr){1-1} \cmidrule(lr){2-7} \cmidrule(lr){8-13}
S\&P 500 (TR) & 25.34 & 18.09 & 1.40 & -1.04 & 2.04 & -23.74 & 22.47 & 20.25 & 1.11 & -0.51 & 1.57 & -22.93 \\
  \cmidrule(lr){1-1} \cmidrule(lr){2-7} \cmidrule(lr){8-13}
ARIMA & 26.74 & 18.09 & 1.48 & -0.05 & 3.39 & -21.90 & 12.62 & 20.25 & 0.62 & -0.34 & 0.91 & -16.82 \\
GP & 31.28 & 18.09 & 1.73 & 0.17 & 3.71 & -25.04 & 10.57 & 20.25 & 0.52 & -0.38 & 0.73 & -16.72 \\
Boost & 32.70 & 18.09 & 1.81 & 0.22 & 4.57 & -26.85 & 5.64 & 20.25 & 0.28 & -0.58 & 0.41 & -19.44 \\
  \cmidrule(lr){1-1} \cmidrule(lr){2-7} \cmidrule(lr){8-13}
PW & 30.50 & 18.09 & 1.69 & 0.13 & 3.69 & -26.26 & 15.04 & 20.25 & 0.74 & -0.24 & 1.06 & -15.90 \\
CU & 28.95 & 18.09 & 1.60 & 0.05 & 3.75 & -23.06 & 29.10 & 20.25 & 1.44 & 0.22 & 2.41 & -27.24 \\
CI & 23.15 & 18.09 & 1.28 & -0.22 & 2.61 & -18.85 & 17.15 & 20.25 & 0.85 & -0.18 & 1.24 & -16.74 \\
ICI & 28.57 & 18.09 & 1.58 & 0.04 & 3.58 & -22.90 & 26.14 & 20.25 & 1.29 & 0.12 & 2.20 & -24.05 \\
  \cmidrule(lr){1-1} \cmidrule(lr){2-7} \cmidrule(lr){8-13}
  \end{tabular}}
  

\makebox[\textwidth]{\begin{tabular}{*{13}c}

&  \multicolumn{6}{c}{1999} & \multicolumn{6}{c}{2000}\\
  \cmidrule(lr){1-1} \cmidrule(lr){2-7} \cmidrule(lr){8-13}
  Portfolio method & Cuml Ret & Ret Vol  & Sharpe & IR & Sortino & Max DD & Cuml Ret & Ret Vol  & Sharpe & IR & Sortino & Max DD \\  
  \cmidrule(lr){1-1} \cmidrule(lr){2-7} \cmidrule(lr){8-13}
S\&P 500 (TR) & 16.15 & 18.04 & 0.90 & -3.27 & 1.34 & -14.82 & -12.80 & 22.18 & -0.58 & -0.50 & -0.81 & -20.07 \\
  \cmidrule(lr){1-1} \cmidrule(lr){2-7} \cmidrule(lr){8-13}
ARIMA & -28.79 & 18.05 & -1.60 & -1.72 & -1.71 & -27.22 & -5.65 & 22.17 & -0.25 & -0.10 & -0.33 & -26.38 \\
GP & -21.79 & 18.05 & -1.21 & -1.58 & -1.40 & -22.85 & -8.89 & 22.17 & -0.40 & -0.23 & -0.51 & -26.27 \\
Boost & 19.04 & 18.05 & 1.06 & -0.12 & 1.64 & -25.24 & -29.64 & 22.17 & -1.34 & -0.83 & -1.55 & -38.33 \\
  \cmidrule(lr){1-1} \cmidrule(lr){2-7} \cmidrule(lr){8-13}
PW & -9.48 & 18.05 & -0.53 & -1.13 & -0.68 & -20.84 & -10.81 & 22.17 & -0.49 & -0.32 & -0.64 & -27.58 \\
CU & -3.66 & 18.05 & -0.20 & -0.94 & -0.27 & -21.42 & 1.46 & 22.17 & 0.07 & 0.15 & 0.09 & -24.08 \\
CI & -19.29 & 18.05 & -1.07 & -1.39 & -1.27 & -20.96 & 6.15 & 22.17 & 0.28 & 0.34 & 0.39 & -23.62 \\
ICI & -32.13 & 18.05 & -1.78 & -1.89 & -1.89 & -30.89 & -6.63 & 22.17 & -0.30 & -0.12 & -0.38 & -27.23 \\
  \cmidrule(lr){1-1} \cmidrule(lr){2-7} \cmidrule(lr){8-13}
  \end{tabular}}
\captionof{figure}{Performance metrics, 1993-2000.}
\label{fg:T8}

\end{table}

\begin{table}[ht]
\footnotesize
  \centering
\makebox[\textwidth]{\begin{tabular}{*{13}c}

&  \multicolumn{6}{c}{2001} & \multicolumn{6}{c}{2002}\\
  \cmidrule(lr){1-1} \cmidrule(lr){2-7} \cmidrule(lr){8-13}
  Portfolio method & Cuml Ret & Ret Vol  & Sharpe & IR & Sortino & Max DD & Cuml Ret & Ret Vol  & Sharpe & IR & Sortino & Max DD \\  
  \cmidrule(lr){1-1} \cmidrule(lr){2-7} \cmidrule(lr){8-13}
S\&P 500 (TR) & -14.12 & 21.51 & -0.67 & -2.25 & -0.92 & -30.91 & -23.24 & 25.99 & -0.89 & -1.43 & -1.28 & -33.81 \\
  \cmidrule(lr){1-1} \cmidrule(lr){2-7} \cmidrule(lr){8-13}
ARIMA & 36.09 & 21.51 & 1.68 & 1.72 & 2.57 & -35.22 & 26.07 & 25.99 & 1.00 & 1.49 & 1.53 & -34.82 \\
GP & 36.18 & 21.51 & 1.68 & 1.67 & 2.60 & -34.88 & 33.51 & 25.99 & 1.29 & 1.73 & 2.10 & -37.44 \\
Boost & 25.65 & 21.51 & 1.19 & 1.34 & 1.64 & -30.70 & 31.67 & 25.99 & 1.22 & 1.65 & 1.99 & -36.37 \\
  \cmidrule(lr){1-1} \cmidrule(lr){2-7} \cmidrule(lr){8-13}
PW & 23.71 & 21.51 & 1.10 & 1.28 & 1.48 & -29.85 & 33.17 & 25.99 & 1.28 & 1.68 & 2.09 & -37.85 \\
CU & 21.26 & 21.51 & 0.99 & 1.24 & 1.39 & -26.00 & 4.75 & 25.99 & 0.18 & 0.83 & 0.24 & -30.08 \\
CI & 35.16 & 21.51 & 1.63 & 1.66 & 2.62 & -33.08 & 22.01 & 25.99 & 0.85 & 1.37 & 1.28 & -31.67 \\
ICI & 34.26 & 21.51 & 1.59 & 1.64 & 2.52 & -33.28 & 26.35 & 25.99 & 1.01 & 1.48 & 1.57 & -35.08 \\
  \cmidrule(lr){1-1} \cmidrule(lr){2-7} \cmidrule(lr){8-13}
  \end{tabular}}

\makebox[\textwidth]{\begin{tabular}{*{13}c}

&  \multicolumn{6}{c}{2003} & \multicolumn{6}{c}{2004}\\
  \cmidrule(lr){1-1} \cmidrule(lr){2-7} \cmidrule(lr){8-13}
  Portfolio method & Cuml Ret & Ret Vol  & Sharpe & IR & Sortino & Max DD & Cuml Ret & Ret Vol  & Sharpe & IR & Sortino & Max DD \\  
  \cmidrule(lr){1-1} \cmidrule(lr){2-7} \cmidrule(lr){8-13}
S\&P 500 (TR) & 25.64 & 17.04 & 1.50 & -2.42 & 2.32 & -28.43 & 9.78 & 11.07 & 0.88 & -1.37 & 1.28 & -12.49 \\
  \cmidrule(lr){1-1} \cmidrule(lr){2-7} \cmidrule(lr){8-13}
ARIMA & -23.61 & 17.03 & -1.39 & -2.48 & -1.53 & -23.77 & 14.26 & 11.07 & 1.29 & 0.26 & 2.62 & -15.21 \\
GP & -27.81 & 17.03 & -1.63 & -2.70 & -1.73 & -26.21 & 15.63 & 11.07 & 1.41 & 0.35 & 3.08 & -16.01 \\
Boost & -31.93 & 17.03 & -1.87 & -2.76 & -1.95 & -30.85 & 12.73 & 11.07 & 1.15 & 0.16 & 2.41 & -14.50 \\
  \cmidrule(lr){1-1} \cmidrule(lr){2-7} \cmidrule(lr){8-13}
PW & -31.18 & 17.03 & -1.83 & -2.80 & -1.86 & -29.23 & 14.71 & 11.07 & 1.33 & 0.30 & 3.02 & -14.74 \\
CU & -20.25 & 17.03 & -1.19 & -2.60 & -1.28 & -22.22 & 12.98 & 11.07 & 1.17 & 0.17 & 2.22 & -15.10 \\
CI & -20.35 & 17.03 & -1.19 & -2.32 & -1.36 & -21.51 & 12.22 & 11.07 & 1.10 & 0.13 & 2.18 & -14.11 \\
ICI & -21.49 & 17.03 & -1.26 & -2.39 & -1.40 & -22.16 & 16.73 & 11.07 & 1.51 & 0.43 & 3.21 & -15.57 \\
  \cmidrule(lr){1-1} \cmidrule(lr){2-7} \cmidrule(lr){8-13}
  \end{tabular}}

\makebox[\textwidth]{\begin{tabular}{*{13}c}

&  \multicolumn{6}{c}{2005} & \multicolumn{6}{c}{2006}\\
  \cmidrule(lr){1-1} \cmidrule(lr){2-7} \cmidrule(lr){8-13}
  Portfolio method & Cuml Ret & Ret Vol  & Sharpe & IR & Sortino & Max DD & Cuml Ret & Ret Vol  & Sharpe & IR & Sortino & Max DD \\ 
  \cmidrule(lr){1-1} \cmidrule(lr){2-7} \cmidrule(lr){8-13}
S\&P 500 (TR) & 2.37 & 10.26 & 0.23 & -3.66 & 0.33 & -9.87 & 10.49 & 10.02 & 1.05 & -3.14 & 1.59 & -12.93 \\
  \cmidrule(lr){1-1} \cmidrule(lr){2-7} \cmidrule(lr){8-13}
ARIMA & 3.65 & 10.27 & 0.36 & -0.24 & 0.52 & -9.98 & -3.92 & 10.02 & -0.39 & -1.17 & -0.58 & -15.33 \\
GP & -0.49 & 10.27 & -0.05 & -0.57 & -0.07 & -11.40 & -2.43 & 10.02 & -0.24 & -1.03 & -0.36 & -14.76 \\
Boost & 1.49 & 10.27 & 0.15 & -0.42 & 0.22 & -11.54 & -4.06 & 10.02 & -0.41 & -1.14 & -0.58 & -15.78 \\
  \cmidrule(lr){1-1} \cmidrule(lr){2-7} \cmidrule(lr){8-13}
PW & 0.10 & 10.27 & 0.01 & -0.53 & 0.01 & -12.06 & -6.36 & 10.02 & -0.63 & -1.33 & -0.87 & -16.78 \\
CU & 1.58 & 10.27 & 0.15 & -0.44 & 0.23 & -10.15 & 3.40 & 10.02 & 0.34 & -0.59 & 0.61 & -13.37 \\
CI & 1.62 & 10.27 & 0.16 & -0.41 & 0.22 & -10.65 & -2.07 & 10.02 & -0.21 & -1.02 & -0.32 & -14.31 \\
ICI & 1.23 & 10.27 & 0.12 & -0.45 & 0.17 & -9.78 & 1.52 & 10.02 & 0.15 & -0.74 & 0.27 & -12.77 \\
  \cmidrule(lr){1-1} \cmidrule(lr){2-7} \cmidrule(lr){8-13}
  \end{tabular}}

\makebox[\textwidth]{\begin{tabular}{*{13}c}

&  \multicolumn{6}{c}{2007} & \multicolumn{6}{c}{2008}\\
  \cmidrule(lr){1-1} \cmidrule(lr){2-7} \cmidrule(lr){8-13}
  Portfolio method & Cuml Ret & Ret Vol  & Sharpe & IR & Sortino & Max DD & Cuml Ret & Ret Vol  & Sharpe & IR & Sortino & Max DD \\ 
  \cmidrule(lr){1-1} \cmidrule(lr){2-7} \cmidrule(lr){8-13}
S\&P 500 (TR) & 2.07 & 15.96 & 0.13 & -4.06 & 0.17 & -10.66 & -39.34 & 40.88 & -0.96 & -0.77 & -1.31 & -47.78 \\
  \cmidrule(lr){1-1} \cmidrule(lr){2-7} \cmidrule(lr){8-13}
ARIMA & -1.06 & 15.96 & -0.07 & -0.33 & -0.10 & -12.74 & 17.00 & 40.88 & 0.42 & 1.12 & 0.60 & -42.30 \\
GP & -1.52 & 15.96 & -0.10 & -0.35 & -0.14 & -13.27 & 46.05 & 40.88 & 1.13 & 1.69 & 1.83 & -40.38 \\
Boost & -3.80 & 15.96 & -0.24 & -0.48 & -0.35 & -15.23 & 39.48 & 40.88 & 0.97 & 1.34 & 1.56 & -43.34 \\
  \cmidrule(lr){1-1} \cmidrule(lr){2-7} \cmidrule(lr){8-13}
PW & 1.83 & 15.96 & 0.11 & -0.21 & 0.18 & -14.36 & 56.22 & 40.88 & 1.38 & 1.81 & 2.61 & -46.29 \\
CU & -13.65 & 15.96 & -0.86 & -0.87 & -1.09 & -17.35 & 67.57 & 40.88 & 1.65 & 2.27 & 3.13 & -51.49 \\
CI & -3.35 & 15.96 & -0.21 & -0.44 & -0.31 & -11.66 & 47.33 & 40.88 & 1.16 & 1.79 & 2.16 & -42.41 \\
ICI & -0.99 & 15.96 & -0.06 & -0.31 & -0.08 & -9.66 & 60.76 & 40.88 & 1.49 & 2.33 & 2.91 & -49.16 \\
  \cmidrule(lr){1-1} \cmidrule(lr){2-7} \cmidrule(lr){8-13}
  \end{tabular}}

\captionof{figure}{Performance metrics, 2001-2008.}
\label{fg:T8}

\end{table}

\begin{table}[ht]
\footnotesize
  \centering
\makebox[\textwidth]{\begin{tabular}{*{13}c}

&  \multicolumn{6}{c}{2009} & \multicolumn{6}{c}{2010}\\
  \cmidrule(lr){1-1} \cmidrule(lr){2-7} \cmidrule(lr){8-13}
  Portfolio method & Cuml Ret & Ret Vol  & Sharpe & IR & Sortino & Max DD & Cuml Ret & Ret Vol  & Sharpe & IR & Sortino & Max DD \\  
  \cmidrule(lr){1-1} \cmidrule(lr){2-7} \cmidrule(lr){8-13}
S\&P 500 (TR) & 27.09 & 27.22 & 1.00 & -0.83 & 1.45 & -41.01 & 15.47 & 18.02 & 0.86 & -1.81 & 1.23 & -19.54 \\
  \cmidrule(lr){1-1} \cmidrule(lr){2-7} \cmidrule(lr){8-13}
ARIMA & 17.54 & 27.22 & 0.64 & -0.13 & 1.13 & -21.65 & -41.11 & 18.02 & -2.28 & -2.30 & -2.19 & -37.28 \\
GP & 7.78 & 27.22 & 0.29 & -0.43 & 0.48 & -18.56 & -42.99 & 18.02 & -2.38 & -2.42 & -2.33 & -38.98 \\
Boost & -33.40 & 27.22 & -1.23 & -1.76 & -1.51 & -35.25 & -33.99 & 18.02 & -1.89 & -2.07 & -1.97 & -33.55 \\
  \cmidrule(lr){1-1} \cmidrule(lr){2-7} \cmidrule(lr){8-13}
PW & -9.72 & 27.22 & -0.36 & -1.00 & -0.55 & -19.77 & -36.86 & 18.02 & -2.04 & -2.17 & -2.06 & -34.57 \\ 
CU & 19.15 & 27.22 & 0.70 & -0.08 & 1.30 & -21.64 & -36.30 & 18.02 & -2.01 & -2.14 & -2.03 & -35.08 \\
CI & 25.99 & 27.22 & 0.95 & 0.12 & 1.52 & -31.85 & -37.32 & 18.02 & -2.07 & -2.14 & -1.98 & -34.19 \\
ICI & 27.58 & 27.22 & 1.01 & 0.17 & 1.74 & -27.00 & -38.24 & 18.02 & -2.12 & -2.16 & -1.99 & -33.98 \\
  \cmidrule(lr){1-1} \cmidrule(lr){2-7} \cmidrule(lr){8-13}
  \end{tabular}}

\makebox[\textwidth]{\begin{tabular}{*{13}c}

&  \multicolumn{6}{c}{2011} & \multicolumn{6}{c}{2012}\\
  \cmidrule(lr){1-1} \cmidrule(lr){2-7} \cmidrule(lr){8-13}
  Portfolio method & Cuml Ret & Ret Vol  & Sharpe & IR & Sortino & Max DD & Cuml Ret & Ret Vol  & Sharpe & IR & Sortino & Max DD \\  
  \cmidrule(lr){1-1} \cmidrule(lr){2-7} \cmidrule(lr){8-13}
S\&P 500 (TR) & 4.78 & 23.24 & 0.21 & 0.83 & 0.28 & -18.64 & 15.63 & 12.73 & 1.24 & -0.23 & 1.89 & -14.23 \\
  \cmidrule(lr){1-1} \cmidrule(lr){2-7} \cmidrule(lr){8-13}
ARIMA & -29.76 & 23.24 & -1.28 & -1.08 & -1.35 & -34.76 & 22.54 & 12.73 & 1.77 & 0.46 & 3.23 & -19.67 \\
GP & -29.17 & 23.24 & -1.26 & -1.03 & -1.33 & -33.83 & 29.53 & 12.73 & 2.32 & 0.86 & 5.28 & -25.51 \\
Boost & -8.30 & 23.24 & -0.36 & -0.41 & -0.45 & -18.13 & 28.89 & 12.73 & 2.27 & 0.83 & 5.22 & -25.12 \\
  \cmidrule(lr){1-1} \cmidrule(lr){2-7} \cmidrule(lr){8-13}
PW & -8.41 & 23.24 & -0.36 & -0.42 & -0.48 & -16.08 & 25.43 & 12.73 & 2.00 & 0.65 & 4.23 & -24.06 \\
CU & -15.96 & 23.24 & -0.69 & -0.56 & -0.80 & -29.10 & 15.64 & 12.73 & 1.23 & 0.09 & 1.79 & -15.94 \\
CI & -34.31 & 23.24 & -1.48 & -1.19 & -1.53 & -32.76 & 19.15 & 12.73 & 1.50 & 0.28 & 2.45 & -18.43 \\
ICI & -26.26 & 23.24 & -1.13 & -0.92 & -1.18 & -33.43 & 18.62 & 12.73 & 1.46 & 0.25 & 2.45 & -16.55 \\
  \cmidrule(lr){1-1} \cmidrule(lr){2-7} \cmidrule(lr){8-13}
  \end{tabular}}

\makebox[\textwidth]{\begin{tabular}{*{13}c}

&  \multicolumn{6}{c}{2013} & \multicolumn{6}{c}{2014}\\
  \cmidrule(lr){1-1} \cmidrule(lr){2-7} \cmidrule(lr){8-13}
  Portfolio method & Cuml Ret & Ret Vol  & Sharpe & IR & Sortino & Max DD & Cuml Ret & Ret Vol  & Sharpe & IR & Sortino & Max DD \\  
  \cmidrule(lr){1-1} \cmidrule(lr){2-7} \cmidrule(lr){8-13}
S\&P 500 (TR)  & 28.68 & 11.05 & 2.60 & -1.62 & 3.96 & -22.78 & 13.48 & 11.33 & 1.19 & 1.09 & 1.67 & -18.27 \\
  \cmidrule(lr){1-1} \cmidrule(lr){2-7} \cmidrule(lr){8-13}
ARIMA & 2.61 & 11.05 & 0.24 & -1.74 & 0.34 & -11.82 & 8.11 & 11.33 & 0.72 & -0.35 & 1.74 & -10.11 \\
GP & -4.37 & 11.05 & -0.40 & -2.45 & -0.51 & -16.53 & 7.28 & 11.33 & 0.64 & -0.42 & 1.33 & -8.94 \\
Boost & 7.30 & 11.05 & 0.66 & -1.47 & 0.96 & -8.04 & 16.40 & 11.33 & 1.45 & 0.18 & 3.85 & -15.84 \\
  \cmidrule(lr){1-1} \cmidrule(lr){2-7} \cmidrule(lr){8-13}
PW & -1.12 & 11.05 & -0.10 & -2.36 & -0.14 & -14.72 & 14.34 & 11.33 & 1.27 & 0.05 & 3.14 & -14.43 \\
CU & -11.24 & 11.05 & -1.02 & -2.96 & -1.17 & -19.21 & 1.81 & 11.33 & 0.16 & -0.79 & 0.29 & -7.87 \\
CI & 4.64 & 11.05 & 0.42 & -1.57 & 0.63 & -9.25 & 6.04 & 11.33 & 0.53 & -0.47 & 1.26 & -9.70 \\
ICI & -3.97 & 11.05 & -0.36 & -2.40 & -0.47 & -15.92 & 7.34 & 11.33 & 0.65 & -0.39 & 1.43 & -9.50 \\
  \cmidrule(lr){1-1} \cmidrule(lr){2-7} \cmidrule(lr){8-13}
  \end{tabular}}
  
\makebox[\textwidth]{\begin{tabular}{*{13}c}

&  \multicolumn{6}{c}{2015} & \multicolumn{6}{c}{2016}\\
  \cmidrule(lr){1-1} \cmidrule(lr){2-7} \cmidrule(lr){8-13}
  Portfolio method & Cuml Ret & Ret Vol  & Sharpe & IR & Sortino & Max DD & Cuml Ret & Ret Vol  & Sharpe & IR & Sortino & Max DD \\  
  \cmidrule(lr){1-1} \cmidrule(lr){2-7} \cmidrule(lr){8-13}
S\&P 500 (TR) & 2.57 & 15.46 & 0.17 & 0.91 & 0.23 & -12.04  & 11.94 & 13.07 & 0.91 & -1.32 & 1.30 & -20.80 \\
  \cmidrule(lr){1-1} \cmidrule(lr){2-7} \cmidrule(lr){8-13}
ARIMA & 29.50 & 15.46 & 1.91 & 1.56 & 2.82 & -24.32 & -0.11 & 13.07 & -0.01 & -0.90 & -0.01 & -11.81 \\
GP & 25.13 & 15.46 & 1.62 & 1.34 & 2.52 & -24.29 & 0.75 & 13.07 & 0.06 & -0.84 & 0.11 & -12.31 \\
Boost & 28.98 & 15.46 & 1.87 & 1.47 & 3.17 & -26.31 & -3.25 & 13.07 & -0.25 & -1.08 & -0.39 & -10.62 \\
  \cmidrule(lr){1-1} \cmidrule(lr){2-7} \cmidrule(lr){8-13}
PW & 28.45 & 15.46 & 1.84 & 1.56 & 2.77 & -26.20 & -11.84 & 13.07 & -0.91 & -1.60 & -1.22 & -12.92 \\
CU & 31.35 & 15.46 & 2.03 & 1.83 & 2.96 & -26.59 & -1.96 & 13.07 & -0.15 & -1.01 & -0.25 & -12.55 \\
CI & 27.01 & 15.46 & 1.75 & 1.45 & 2.56 & -22.93 & 2.62 & 13.07 & 0.20 & -0.73 & 0.38 & -13.80 \\
ICI & 22.50 & 15.46 & 1.46 & 1.12 & 2.42 & -23.22 & 6.23 & 13.07 & 0.48 & -0.52 & 1.03 & -13.95 \\
  \cmidrule(lr){1-1} \cmidrule(lr){2-7} \cmidrule(lr){8-13}
  \end{tabular}}

\captionof{figure}{Performance metrics, 2009-2016.}
\label{fg:T8}
\end{table}

\begin{table}[ht]
\footnotesize
  \centering
\makebox[\textwidth]{\begin{tabular}{*{13}c}

&  \multicolumn{6}{c}{2017} & \multicolumn{6}{c}{2018}\\
  \cmidrule(lr){1-1} \cmidrule(lr){2-7} \cmidrule(lr){8-13}
  Portfolio method & Cuml Ret & Ret Vol  & Sharpe & IR & Sortino & Max DD & Cuml Ret & Ret Vol  & Sharpe & IR & Sortino & Max DD \\  
  \cmidrule(lr){1-1} \cmidrule(lr){2-7} \cmidrule(lr){8-13}
S\&P 500 (TR) & 19.16 & 6.67 & 2.89 & -1.04 & 4.55 & -17.04 & -4.81 & 17.03 & -0.28 & -0.77 & -0.37 & -19.81 \\
  \cmidrule(lr){1-1} \cmidrule(lr){2-7} \cmidrule(lr){8-13}
ARIMA & -1.08 & 6.67 & -0.16 & -2.14 & -0.24 & -7.76 & 4.96 & 17.03 & 0.29 & 0.50 & 0.40 & -14.97 \\
GP & 0.06 & 6.67 & 0.01 & -2.03 & 0.02 & -5.84 & -0.59 & 17.03 & -0.03 & 0.26 & -0.04 & -14.22 \\
Boost & -3.86 & 6.67 & -0.58 & -2.50 & -0.79 & -10.74 & 6.59 & 17.03 & 0.39 & 0.61 & 0.56 & -16.52 \\
  \cmidrule(lr){1-1} \cmidrule(lr){2-7} \cmidrule(lr){8-13}
PW & -7.71 & 6.67 & -1.16 & -2.99 & -1.41 & -12.44 & 13.39 & 17.03 & 0.79 & 0.87 & 1.17 & -17.03 \\
CU & -0.88 & 6.67 & -0.13 & -2.16 & -0.20 & -6.45 & 7.20 & 17.03 & 0.42 & 0.57 & 0.57 & -12.90 \\
CI & 2.51 & 6.67 & 0.38 & -1.75 & 0.61 & -6.02 & 2.18 & 17.03 & 0.13 & 0.37 & 0.17 & -12.58 \\
ICI & 4.71 & 6.67 & 0.71 & -1.48 & 1.32 & -4.78 & -5.87 & 17.03 & -0.34 & 0.05 & -0.44 & -14.05 \\
  \cmidrule(lr){1-1} \cmidrule(lr){2-7} \cmidrule(lr){8-13}
  \end{tabular}}
  
  \vspace{0.5 \baselineskip}

\makebox[\textwidth]{\begin{tabular}{*{13}c}

&  \multicolumn{6}{c}{2019} & \multicolumn{6}{c}{2020}\\
  \cmidrule(lr){1-1} \cmidrule(lr){2-7} \cmidrule(lr){8-13}
  Portfolio method & Cuml Ret & Ret Vol  & Sharpe & IR & Sortino & Max DD & Cuml Ret & Ret Vol  & Sharpe & IR & Sortino & Max DD \\  
  \cmidrule(lr){1-1} \cmidrule(lr){2-7} \cmidrule(lr){8-13}
S\&P 500 (TR) & 26.03 & 12.45 & 2.09 & -0.79 & 3.02 & -24.34 & 22.51 & 34.37 & 0.65 & -2.26 & 0.90 & -41.22 \\
  \cmidrule(lr){1-1} \cmidrule(lr){2-7} \cmidrule(lr){8-13}
ARIMA & 22.69 & 12.45 & 1.37 & -0.16 & 3.17 & -22.74 & 17.36 & 34.37 & 0.51 & -0.17 & 0.61 & -38.09 \\
GP & 20.72 & 12.45 & 1.66 & -0.30 & 3.69 & -21.68 & -1.95 & 34.37 & -0.06 & -0.59 & -0.07 & -36.61 \\
Boost & 23.24 & 12.45 & 1.87 & -0.15 & 4.15 & -20.27 & 26.22 & 34.37 & 0.76 & 0.03 & 1.01 & -39.32 \\
  \cmidrule(lr){1-1} \cmidrule(lr){2-7} \cmidrule(lr){8-13}
PW & 16.83 & 12.45 & 1.35 & -0.50 & 3.33 & -19.62 & 22.99 & 34.37 & 0.67 & -0.04 & 0.86 & -37.70 \\
CU & 17.75 & 12.45 & 1.43 & -0.50 & 2.73 & -19.82 & 3.84 & 34.37 & 0.11 & -0.47 & 0.13 & -36.68 \\
CI & 23.50 & 12.45 & 1.89 & -0.15 & 3.68 & -21.02 & 8.55 & 34.37 & 0.25 & -0.36 & 0.29 & -36.89 \\
ICI & 19.15 & 12.45 & 1.54 & -0.42 & 2.47 & -19.28  & 8.74 & 34.37 & 0.25 & -0.36 & 0.30 & -36.94 \\
  \cmidrule(lr){1-1} \cmidrule(lr){2-7} \cmidrule(lr){8-13}
  \end{tabular}}

\vspace{0.5 \baselineskip}

\makebox[\textwidth]{\begin{tabular}{*{13}c}

&  \multicolumn{6}{c}{2021} & \multicolumn{6}{c}{ }\\
  \cmidrule(lr){1-1} \cmidrule(lr){2-7} \cmidrule(lr){8-13}
  Portfolio method & Cuml Ret & Ret Vol  & Sharpe & IR & Sortino & Max DD \\  
  \cmidrule(lr){1-1} \cmidrule(lr){2-7} \cmidrule(lr){8-13}
S\&P 500 (TR) & 26.10 & 13.07 & 2.00 & 1.55 & 2.97 & -23.86 \\
  \cmidrule(lr){1-1} \cmidrule(lr){2-7} \cmidrule(lr){8-13}
ARIMA & -5.78 & 13.07 & -0.46 & -1.65 & -0.60 & -12.93 \\
GP & -2.48 & 13.07 & -0.20 & -1.47 & -0.27 & -10.21 \\
Boost & 6.64 & 13.07 & 0.53 & -0.81 & 0.84 & -12.56 \\
  \cmidrule(lr){1-1} \cmidrule(lr){2-7} \cmidrule(lr){8-13}
PW & -5.24 & 13.07 & -0.42 & -1.55 & -0.53 & -12.61 \\
CU & -0.34 & 13.07 & -0.03 & -1.37 & -0.04 & -9.50 \\
CI & -1.12 & 13.07 & -0.09 & -1.34 & -0.13 & -9.80 \\
ICI & -2.15 & 13.07 & -0.17 & -1.40 & -0.24 & -10.50 \\
  \cmidrule(lr){1-1} \cmidrule(lr){2-7} \cmidrule(lr){8-13}
  \end{tabular}}

\captionof{figure}{Performance metrics, 2017-2021.}
\label{fg:T8}
\end{table}

\end{document}